\documentclass{article}
\usepackage{authblk}
\usepackage{tikz}
\usepackage[english]{babel}
\usepackage[letterpaper,top=2cm,bottom=2cm,left=3cm,right=3cm,marginparwidth=1.75cm]{geometry}
\usepackage{amsmath}
\usepackage{amsfonts}
\usepackage{amssymb}
\usepackage{array}
\usepackage{booktabs}
\usepackage{graphicx}
\usepackage[colorlinks=true, allcolors=blue]{hyperref}
\usepackage{tikz}
\usetikzlibrary{external}
\usepackage{multirow}
\usepackage{pdfpages}
\usetikzlibrary{positioning}
\usepackage{verbatim}
\usetikzlibrary{arrows,shapes, arrows.meta}
\usepackage{hyperref}
\usepackage{nameref}
\usepackage{enumitem}
\usepackage{caption}
\usepackage{graphicx}
\usepackage{multirow}
\usepackage{setspace}
\usepackage{subcaption}
\usepackage{relsize}
\usepackage{xcolor, soul}
\usepackage{verbatim}
\usepackage{ntheorem}
\usepackage{algorithm}
\usepackage{algpseudocode}
\usepackage{xurl}

\theoremstyle{plain}

\theoremstyle{plain}

\theoremstyle{plain}

\theoremstyle{plain}

\title{Accurate stochastic simulation algorithm for multiscale models of infectious diseases}
\author[1]{Yuan Yin}
\author[2]{Jennifer A. Flegg}
\author[3]{Mark B. Flegg}
\affil[1]{Mathematical Institute, The University of Oxford}
\affil[2]{School of Mathematics and Statistics, The University of Melbourne}
\affil[3]{School of Mathematics, Monash University}

\date{}

\begin{document} 
\maketitle

\begin{abstract}
In the infectious disease literature, significant effort has been devoted to studying dynamics at a single scale.~For example, compartmental models describing population-level dynamics are often formulated using differential equations.~In cases where small numbers or noise play a crucial role, these differential equations are replaced with memoryless Markovian models, where discrete individuals can be members of a compartment and transition stochastically.~Classic stochastic simulation algorithms, such as the next reaction method, can be employed to solve these Markovian models exactly.~The intricate coupling between models at different scales underscores the importance of multiscale modelling in infectious diseases.~However, several computational challenges arise when the multiscale model becomes non-Markovian.~In this paper, we address these challenges by developing a novel exact stochastic simulation algorithm.~We apply it to a showcase multiscale system where all individuals share the same deterministic within-host model while the population-level dynamics are governed by a stochastic formulation.~We demonstrate that as long as the within-host information is harvested at a reasonable resolution, the novel algorithm will always be accurate.~Furthermore, our implementation is still efficient even at finer resolutions.~Beyond infectious disease modelling, the algorithm is widely applicable to other multiscale systems, providing a versatile, accurate, and computationally efficient framework.
\end{abstract}

\section{Introduction}
Infectious diseases exhibit complex dynamics and are governed by various spatial and temporal scales, which may include within-host infection processes, host-vector interactions, and between-host transmission patterns \cite{almocera2018multiscale}.~These scales interact with each other; for example, two infectious individuals carrying the same disease can have significantly different infectivity levels in transmitting the disease at the population scale.~Often these differences are important because of the way a disease undergoes a branching effect in a population \cite{smith2005hiv}, such as super-spreaders.~Due to the complicated nature of infectious diseases, one vital tool to investigate and provide understanding of disease dynamics is mathematical modelling \cite{heesterbeek2015modeling}.\\ 
 
The importance of mathematical tools for informing response to infectious diseases has been made abundantly clear by the COVID-19 pandemic.~Through mathematical modelling, transmission pathways, common patterns, and the underlying mechanisms which govern disease dynamics can be explored and analysed \cite{diekmann2012mathematical}.~Additionally, in situations where experiments can be unethical or direct observation is infeasible, mathematical modelling can be applied to estimate and predict the future dynamics of infectious diseases.~However, there remains a pressing need for theoretical and computational innovation in modelling techniques across all spatio-temporal and population scales, and even more so for models that integrate these scales.\\

Single scale epidemiological models have a limited scope for exploring a wide range of research questions.~Disease population dynamics are most often modelled at macroscopic scales using ordinary differential equations (ODEs) to represent distinct populations (susceptible, infected, etc.) and their changes over time (due to infections, death, vaccination etc.).~Deterministic trajectories of the disease can be computed and finding basins of attraction, fixed point stability, and basic reproduction numbers, $R_0$, give rapid insight into the fundamental macroscopic behaviour of the disease \cite{keeling2008modeling}.~However, such models do not capture stochasticity or individual critical events and therefore cannot be used to quantify uncertainty of predictions, find probabilities of elimination, or understand the early stages of a disease. \\

Multiscale models, which combine methods at different scales, are ubiquitous in scientific applications \cite{garabed2020multi, garira2018primer, garira2020research, garira2019coupled, garira2014mathematical}.~In particular, whole-cell simulation is one of the `Grand Challenges of the 21st Century' \cite{tomita2001whole}.~Researchers in this field aim to develop complete models of biological cells as autonomous molecular machines.~Models that fail to appropriately account for the scale of the problem at hand (for example, using a macroscopic model for small population sizes or using a microscopic model for large populations) will either lead to unreliable results or unacceptable computational times.~In times of urgency, such as the emergence of a new infectious pathogen, this causes delays and inaccuracies in generating scenario exploration, hypothesis tests and quantitative predictions.~Instead, there is a need for sophisticated modelling tools that are general enough to handle an array of modelling approaches at scales most appropriate to provide accurate insights and therefore contribute more effectively to informed decision making related to the threat of infectious diseases.  \\ 

There is a pressing need for more research to accurately and efficiently simulate stochastic models which capture the dynamics of each individual embedded in large ensemble populations.~At small population scales, including sub-populations like the population of infected individuals, the state of individuals is critically important, and often a Stochastic Simulation Algorithm (SSA) will be used.~In an SSA, individual events, such as infection, or recovery, of a specific individual, are simulated.~To achieve this, event propensities (rates) $\alpha$ for all possible events are calculated, and then wait times for these possible events are sampled from probability densities (exponentially distributed with rate $\alpha$).~The event that occurs first is then adopted as part of the simulation, and the time until this event $\tau$ is used to update the state of the simulation.~This process is repeated for as long as necessary, noting that each time the state of the simulation is updated, the future propensities for some events $\alpha$ may change.~For example, if an event creates a new infected individual, the propensity for new infections increases proportionally. \\

Exact and approximate SSAs are derived from Gillespie’s foundational work on models of molecular biology \cite{gillespie1977exact}.~For instance, variants such as the Next Reaction Method \cite{gibson2000efficient} make clever use of the memorylessness of exponentially-distributed wait times to improve simulation times.~On the other hand, approximate methods such as Tau-Leaping \cite{cao2006efficient} allow for significant computational savings by implementing a bundle of events over a discrete time interval all at once.~It forgoes updating the system state after every individual event, under the assumptions that (1) each individual event has only a small effect on propensities and (2) the changing of propensities over time can be approximated by state changes due to the cumulative effect of a bundle of events.~With a few exceptions, SSAs are almost always applied under the Markovian assumption that the event propensities $\alpha$ is (piecewise) constant between updates and $\tau$ is an exponentially-distributed wait time.~However, in the case of infectious diseases, this is often not appropriate.~For example, in situations where it is necessary to couple rapid events (such as events associated with within-host models) and relatively slow events (such as population-level changes), it becomes too computationally intensive to simulate all of the within-host events for each host in a population whilst simultaneously accounting for the evolution of behaviour at the population level in response to within-host temporal dynamics for each individual.~Those events, in addition, are not Markovian in nature, so that event propensities $\alpha$ become functions of system states and time and the distribution of $\tau$ is much more complex.~Hence, there is a need for innovative methods which allow for accurate simulation in cases that involve multiple scales.~We investigate this problem by developing an exact simulation framework for multiscale infectious disease models that incorporate non-Markovian dynamics.\\

Previous efforts have been made under the assumption that the distribution of event wait times, $\tau$, is known, analytic, and fixed.~For example, an approximate algorithm based on series expansion \cite{boguna2014simulating}, exact algorithms using the Laplace transform \cite{masuda2018gillespie} and based on next reaction method \cite{gibson2000efficient}, and an algorithm for non-Markovian processes on time-varying networks \cite{vestergaard2015temporal} have been proposed.~In contrast, our work relaxes this assumption, allowing $\tau$ to follow an arbitrary time-varying distribution as defined by the multiscale infectious disease model.~Moreover, our work incorporates individuality by considering how each individual’s within-host system influences $\tau$.~Researchers have also employed a second-order series expansion approximation to avoid the computational expense of numerically integrating the propensity function $\alpha$ in non-Markovian settings \cite{carletti2012stochastic}.~This method assumes (1) $\alpha$ and its derivatives have well-defined analytic forms at the population level, and (2) the time variation in $\alpha$ is sufficiently slow for a second-order approximation to be accurate.~However, these assumptions may not hold in multiscale infectious disease systems.~For example, if disease transmission depends on each individual’s within-host viral load, then the transmission propensity $\alpha$ will vary for each individual, and the functional form of $\alpha$ may not be analytically expressed.~Additionally, if the within-host dynamics evolve rapidly, a second-order approximation may not suffice.~Our work addresses these limitations by relaxing these constraints, thereby enhancing the generality of our framework.\\

Another relevant algorithm, the `modified next reaction method', was introduced by Anderson \cite{anderson2007modified}.~In this approach, the wait time $\tau$ for an event is calculated by taking the antiderivative of the propensity function $\alpha$, which depends on both the system state and time, and then performing inversions.~The primary distinction between this method and our novel framework lies in the definition of $\alpha$.~Instead of defining $\alpha$ as a function that varies with the system state and time, we define $\alpha$ to be state-invariant by leveraging information about how infectious disease models at different scales couple to one another at the level of each infected individual.~This formulation enables us to compute the integrated propensity function only once for the entire simulation.~We then store the computed values in look-up tables for efficient inversion, incorporating resampling and rescaling of the wait time $\tau$.~Hence, another key contribution of our work is its enhanced practicality.\\

The paper is structured as follows.~In Section \ref{methods}, we develop the multiscale model and present the simulation algorithm.~The simulation algorithm is the key innovation of this paper, as it exploits common features among individuals to deliver accurate and efficient results.~The innovation of the algorithm allows for its adoption in a broad range of multiscale infectious disease models.~Then in Section \ref{results}, we present simulation results for some key test problems that highlight the utility of the new algorithm.~We show that our algorithm achieves superior accuracy without sacrificing efficiency.~Finally, in Section \ref{discussion}, we discuss the implications of our new framework and simulation algorithm in a broader context.

\section{Methods}\label{methods}

\subsection{Model framework}

\subsubsection{Population-level model}
\label{population level model}
The simplest disease model, which underpins the core of most infectious disease models, describes the population-level dynamics of transmission between those susceptible to a disease and those capable of infecting others \cite{feng2012model, murillo2013towards}.~To demonstrate the multiscale framework of this paper, we have opted to keep the disease model at the population scale as simple as possible.~By assuming that individuals are removed at a constant rate in both susceptible and infectious states and that new individuals are introduced as susceptible, we adopt the well-known deterministic $S$ (Susceptible) -- $I$ (Infectious) ODE model \cite{anderson1992infectious}: 
\begin{align}
\label{BH submodel}
    \begin{split}
        \frac{dS}{dt} &=\alpha^{(b)} - \alpha^{(i)}(S,I) - \alpha^{(ds)}(S) = \Lambda - \beta SI - \mu S, \quad S(0) = S_0;\\
        \frac{dI}{dt} &= \alpha^{(i)}(S,I) - \alpha^{(di)}(I) = \beta SI - \mu I, \quad I(0) = I_0, \\
    \end{split}
\end{align}
where $S$ and $I$ denote the number of susceptible and infectious individuals, respectively.~The terms $\alpha^{(b)} = \Lambda$, $\alpha^{(ds)}(s) = \mu s$, $\alpha^{(di)}(i) = \mu i$, and $\alpha^{(i)}(s,i) = \beta si$ represent the rates of each of the events modelled within the system.~These events are as follows:~birth of susceptible, death of susceptible, death of infected and infection of susceptible individuals, respectively.~Importantly, we use $\alpha$ in general to represent rates and superscripts $(b)$, $(ds)$, $(di)$ and $(i)$ to indicate specificity to each of the four event types.~The constants $\Lambda$ and $\mu$ are, respectively, the rate at which new susceptible individuals enter the system, and the per capita death rate.~The disease transmission coefficient is given by $\beta$.~The schematic diagram for this population-level model is shown in \autoref{BH_Schematic}.\\

Model (\ref{BH submodel}) encompasses fundamental features of most population scale infectious disease models.~It, therefore, has been extensively studied as a cornerstone base case.~We will extend this model to include multiscale behaviour in two steps: 
\begin{enumerate} 
\item \textit{Use a stochastic simulation algorithm (SSA) at the population level}.~We note that as an ODE, the model only describes large continuous populations, and when populations $S$ or $I$ are small, the model is inappropriate (see Section \ref{SI:linear stability} in Supplementary Information).~We will demonstrate in Section \ref{chap:SSA} how to extend model (\ref{BH submodel}) into a stochastic simulation algorithm.~Although for model (\ref{BH submodel}), such conversion is widely understood, we shall highlight particular ideas useful for generating a multiscale model.
\item \textit{Introduce a deterministic within-host model for individual infections}.~We note that one big assumption in model (\ref{BH submodel}) is that the rates (or propensities) for population-level events to take place do not evolve during the inter-event time intervals.~However, such an assumption may be an oversimplification in a real-world scenario.~A variety of experimental observations \cite{edenborough2012mouse, blaser2014impact} indicate that temporal changes in an infectious agent's within-host viral load can be associated with its propensity to infect others.~We, therefore, extend our model (\ref{BH submodel}) by introducing a deterministic within-host model for the infectious population.~In Section \ref{chap:SSA}, we will highlight how time-dependent event propensities complicate the SSA algorithm.~Then in Section \ref{chap:WH}, we will present a simple deterministic within-host model. 
\end{enumerate}
In Section \ref{MultiscaleSSA}, we combine the population-level SSA model and the within-host model to obtain a multiscale framework.~The aim is to demonstrate an effective way to couple the deterministic changes of individual viral loads with stochastic simulations at the population level.~Whilst we explore this with a simple dynamic model at both scales, the ideas in Section \ref{MultiscaleSSA} are applicable to other multiscale models in a more general setting.

\subsubsection{Converting population-level model into an SSA}
\label{chap:SSA}
The underlying discrete stochastic process of population-level model (\ref{BH submodel}) is a continuous time Markov chain  (CTMC), which describes the evolution of the discrete stochastic non-negative state $(\mathcal{S}(t),\mathcal{I}(t))\in \mathbb{N}_{\geq 0}^2$. $\mathcal{S}(t)$ and $\mathcal{I}(t)$, respectively, represent the number of susceptible and infectious individuals at time $t \in [0, \infty)$.~They are piece-wise constant functions in time because the system remains in a constant state for an exponentially distributed amount of time until moving to the next state \cite{allen2010introduction}. Moreover, a CTMC has the Markovian property of being memory-less;  its future state only depends on the current state of the system
\cite{allen2017primer}. \\

Consider an arbitrary time $t$ when $\mathcal{S}(t) = s$ and $\mathcal{I}(t) = i$, we then define transition probabilities of the stochastic process from state $(s, i)$ to state $(s + m, i + n)$ by time $t + \Delta t$ as
\begin{align}
    \label{Def of transition prob}
    \begin{split}
         &P_{(s, i), (s + m, i + n)}(\Delta t)\\
         &:=\mathbb{P}\Big(\left(\mathcal{S}(t + \Delta t), \mathcal{I}(t + \Delta t) \right) =(s + m, i + n) \big|\left(\mathcal{S}(t), \mathcal{I}(t) \right) = (s, i)\Big),
    \end{split}
\end{align}
where $\Delta t > 0$ is any positive interval of time.
If the initial state of the system $(\mathcal{S}(0),\mathcal{I}(0))=(\mathcal{S}_0,\mathcal{I}_0)$ is known (or subject to a known distribution), then using the notation $P_{(s_0, i_0), (s, i)}(t) := P_{(s, i)}(t)$, the forward Kolmogorov equation to describe the stochastic system associated with the population-level ODE model (\ref{BH submodel}) can be written as:
\begin{align}
    \label{forward Kolmogorov differential equation for BH submodel}
    \begin{split}
        \frac{dP_{(s, i)}}{dt}
        = &\alpha^{(b)} P_{(s-1, i)}(t) + \alpha^{(ds)}(s+1)P_{(s+1, i)}(t) + \alpha^{(di)}P_{(s, i+1)}(t) \\
        &+ \alpha^{(i)}(s+1,i-1) P_{(s+1, i - 1)}(t) - P_{(s, i)}(t)\big[ \alpha^{(b)} + \alpha^{(ds)}(s) + \alpha^{(di)}(i) + \alpha^{(i)}(s,i) \big].
    \end{split}
\end{align}
Where possible, we shall reserve $t$ to represent `calendar time'; the time that has passed since the start of the population-level model.\\

The stochastic simulations of (\ref{forward Kolmogorov differential equation for BH submodel}) -- exact or approximate -- are well-known in the literature and can be achieved in a number of ways.~The simplest approximate approach uses a small set time step $\Delta t$, thus it is referred to as a time-driven algorithm.~At each moment in time of a simulation, the current state of the system $(\mathcal{S}(t),\mathcal{I}(t))=(s,i)$ is known.~Then, multiplying (\ref{forward Kolmogorov differential equation for BH submodel}) by $\Delta t$ gives the update formula to the leading order:

 \begin{align}
    \label{update eq} P_{(s,i)}(t + \Delta t) &= 1 - \left(\alpha^{(b)} + \alpha^{(ds)}(s) + \alpha^{(di)}(i) + \alpha^{(i)}(s,i)\right) \Delta t = 1 - \alpha(s,i) \Delta t; \\
     \label{update eq1}
     P_{(s+1,i)}(t + \Delta t) &= \alpha^{(b)} \Delta t; \\
     P_{(s-1,i)}(t + \Delta t) &= \alpha^{(ds)}(s) \Delta t; \\
     P_{(s,i-1)}(t + \Delta t) &= \alpha^{(di)}(i) \Delta t ;\\
      \label{update eq4}
     P_{(s-1,i+1)}(t + \Delta t) &= \alpha^{(i)}(s,i) \Delta t,
 \end{align}
 where $\alpha$ (with no superscript) is the sum of all (current) event rates. In stochastic contexts, the rates which we denote here using $\alpha$'s are called propensities. We say that $\alpha$ is the propensity for any event to occur. It follows that the probability of this event being birth, susceptible death, infectious death or infection is given by $\alpha^{(b)}/\alpha$, $\alpha^{(ds)}(s)/\alpha$, $\alpha^{(di)}(i)/\alpha$ and $\alpha^{(i)}(s,i)/\alpha$ respectively. The probability of any one of these events taking place in a given time step according to (\ref{update eq}) is $\alpha(s,i)\Delta t$. For numerical simulation of the system, draw a uniform random number from 0 to 1.~If it is less than $\alpha(s,i)\Delta t$, then $(\mathcal{S}(t),\mathcal{I}(t))$ is changed according to the event chosen from the weighted distribution of all possible events.~It is straight-forward to see that this algorithm requires a very small $\Delta t$ for accuracy, which can lead to no events taking place in most time steps.\\

We then introduce the idea underlying the exact Next Reaction Method (NRM), as we will extend and apply a similar strategy to a later multiscale system in Section \ref{clever}.~We consider the forward Kolmogorov equation and ask the question
\begin{align}
    \label{question}
    \text{`If } \Delta t \rightarrow 0\text{, how long does one need to wait before realising an event?'}.
\end{align}
Using (\ref{update eq}), we can determine a probability density function $\phi_{(s,i)}$ in time for a state change if the current state is $(s,i)$:
\begin{equation}
\label{distributionconstantalpha}
\phi_{(s,i)}(T) = \alpha(s,i) \exp\left(-\alpha(s,i) T \right). 
\end{equation}
The use of the upper case $T$ emphasises that (\ref{distributionconstantalpha}) is a distribution for a wait time from the current time $t$ (i.e.~for a putative calendar time of $t+T$).~The putative (event wait) time, $\tau$ can be drawn from $\tau \sim \phi_{(s, i)}(T)$.~The use of the term `putative' is common.~This is because when $\tau$ is sampled, the event is scheduled for the calendar time $t+\tau$.~Note this schedule may change if something changes the propensity $\alpha(s,i)$.\\

The strategy employed in the NRM to address the question in (\ref{question}) relies on two key properties of the exponential distribution (\ref{distributionconstantalpha}).~First, given any sensible partition of the state propensity by event type, $\alpha = \sum_{(e) \in \mathcal{E}} \alpha^{(e)}$, where $\mathcal{E}$ represents the set of possible events (e.g.,~in this manuscript, $\mathcal{E} = \{(b), (ds), (di), (i)\}$), inverse sampling can be utilised to determine the next event time:
\begin{equation} 
    \tau = \mathrm{min}_{(e)\in\mathcal{E}} \left[\tau^{(e)}\right] = \mathrm{min}_{(e)\in\mathcal{E}} \left[ ( \alpha^{(e)} )^{-1} \log\left( (u^{(e)})^{-1} \right) \right],
    \label{eventsampletime}
\end{equation}
where $u^{(e)}\sim U(0,1)$ is independently sampled for each possible event and $\tau^{(e)}$ is the putative time associated with realising each event type.~The most imminent event, denoted as event $(m)$, is the one with the smallest putative wait time and is chosen to propagate the simulation forward (i.e., the “next reaction”).\\

\begin{figure}[t]
    \centering   
    \begin{subfigure}[b]{0.38\textwidth}
         \centering
         \includegraphics[width=\textwidth]{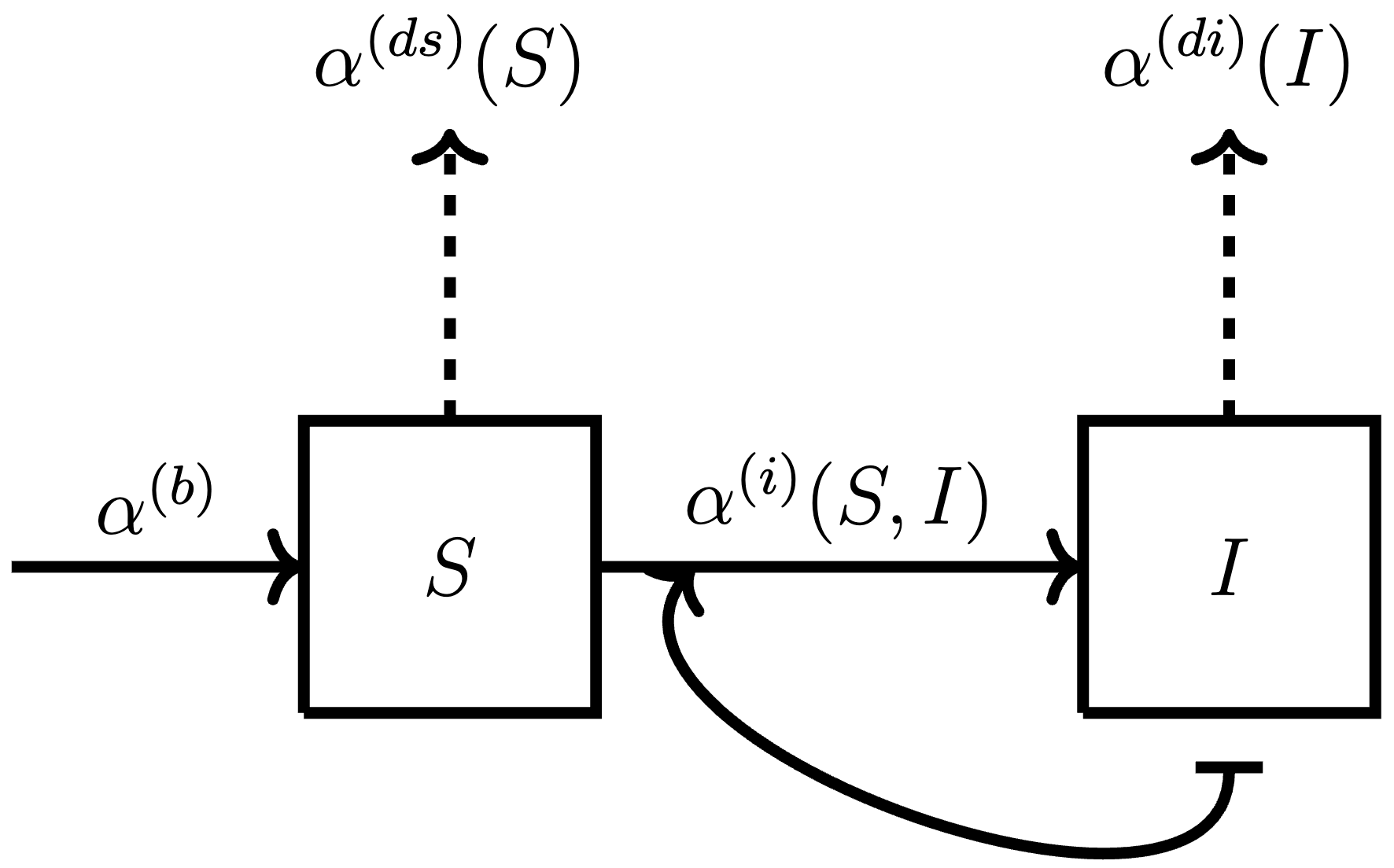}
         \caption{}
         \label{BH_Schematic}
    \end{subfigure}
    \hfill
    \begin{subfigure}[b]{0.61\textwidth}
         \centering
         \includegraphics[width=\textwidth]{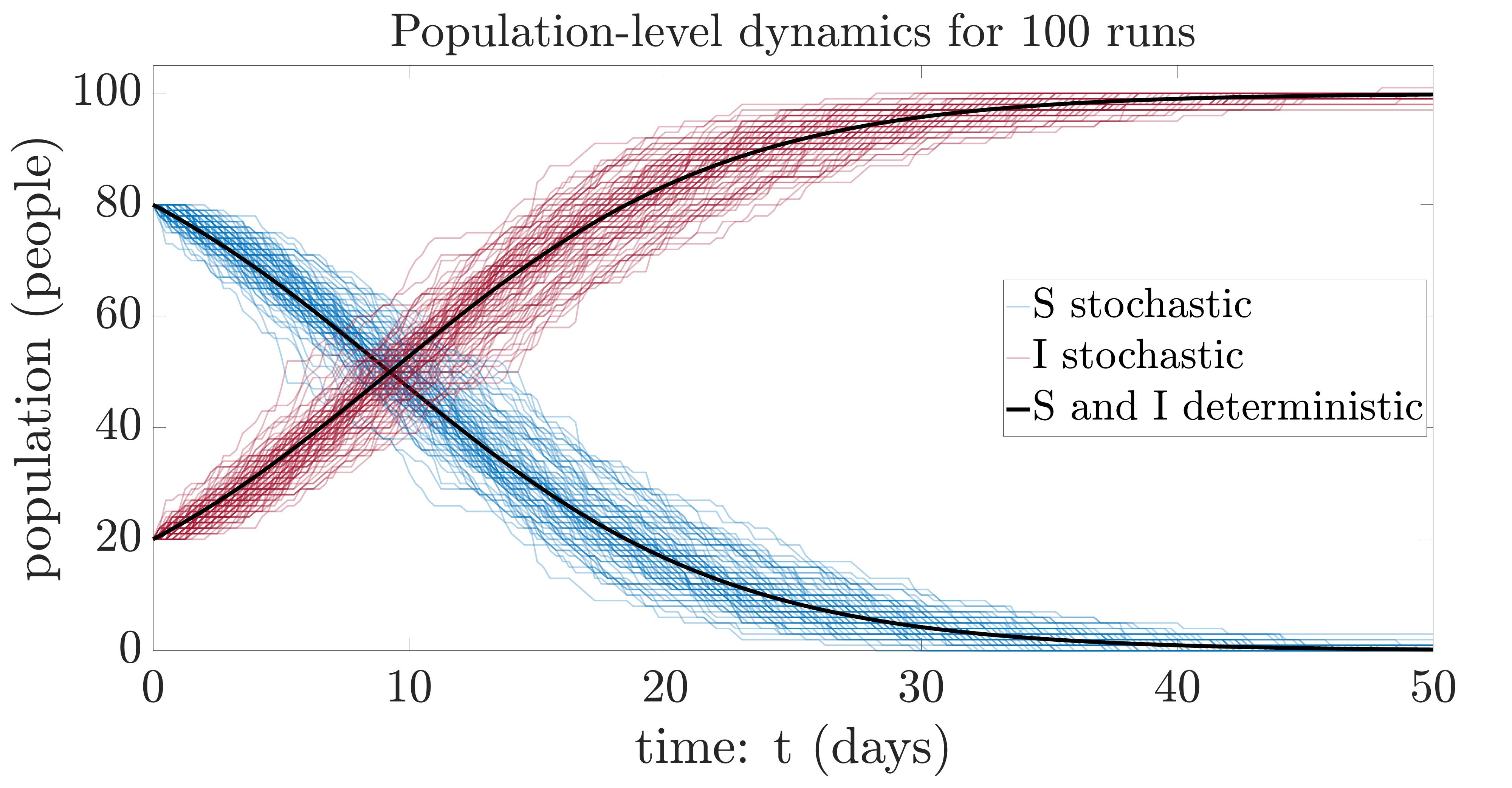}
         \caption{}
         \label{BH_stochastic simulation}
    \end{subfigure}
    \caption{\textbf{(a) Schematic diagram for the population-level model.}~The boxes represent different compartments while the arrows denote transition between compartments.~The compartment linked to the tail end of the arrow, $(\mapsto)$, does not experience any population flow in the corresponding transition.~The dashed arrows denote the removal rates.~\textbf{(b) $\mathbf{100}$ realisations of the population-level model.}~The stochastic simulations are implemented based on the strategy of NRM while the deterministic result is generated by MATLAB’s $ode15s$ solver with both absolute and relative tolerances being $10^{-8}$.~We set the initial condition $\left(S_0, I_0\right)$ as $(80, 20)$ people, and the ending time of the simulation to be $50$ days.~The parameter values chosen are: $\Lambda = 10^{-3}\text{ person}\times \text{day}^{-1}$, $\beta =1.5\times 10^{-3} \rm \text{ }day^{-1}\rm person^{-1}$, and $\mu = 10^{-5}\rm \text{ }day^{-1}$.}
    \label{Sub-system schematic}
\end{figure}

The second property of the exponential distribution (\ref{distributionconstantalpha}) enables the NRM to update putative times when propensities change without discarding and resampling (rescheduling) putative times for all events $(e) \neq (m)$, recall $(m)$ is the most imminent event.~Instead, the times $\tau^{(e)}$ are retained and updated by first subtracting $\tau^{(m)}$ ($\tau^{(e)’} = \tau^{(e)} - \tau^{(m)}$).~We use the dash notation to indicate that a wait time has been updated to a new wait time due to the moving forward of time.~Then the state is changed based on the event $(m)$, which subsequently changes many of the event propensities.~Because each of the putative times $\tau^{(e)}$ have been initially sampled from exponential distributions, their updated values $\tau^{(e)'}$ are conditioned on $\tau^{(e)} > \tau^{(m)}$, which are also exponentially distributed random numbers with propensity $\alpha^{(e)}$ (the lack of markings here highlighting that this is the propensity prior to any state change due to event $(m)$) and there is no need to resample the putative event time.~Thus, a new putative time needs to be generated for the event $(m)$.~Moreover, the putative times for all other events $(e) \neq (m)$ only need to be rescaled following the occurrence of event $(m)$, as shown in (\ref{rescaling}):
\begin{equation} \label{rescaling}
\bar{\tau}^{(e)} = \frac{\alpha^{(e)}}{\bar{\alpha}^{(e)}}\tau^{(e)'} = \frac{\alpha^{(e)}}{\bar{\alpha}^{(e)}} \left( \tau^{(e)}-\tau^{(m)} \right),
\end{equation}
where the bar indicates that the quantity has been updated for the change of current time from $t$ to $t+\tau^{(m)}$ (the dashed notation), and also updated for the change of state/propensity associated with event $(m)$ from the time $t+\tau^{(m)}$ to the new putative calendar event time of $t+\tau^{(m)} + \bar{\tau}$.~After all calculations are done to update the state and putative times for the event $(m)$, $t+\tau^{(m)}$ becomes the current time $t$ and the new state becomes the current state (so bars and dashes are subsequently dropped to be used to denote the updates for the next event).~In \autoref{BH_stochastic simulation} we show simulations of this algorithm against the deterministic ODE model (\ref{BH submodel}).
 \\

The purpose of discussing the strategy used in NRM in detail here is to highlight the importance of the Markovian property in the population-level system.~That is, event times are exponentially distributed (\ref{distributionconstantalpha}) and do not depend on history or time explicitly between events (when the state of the system is constant).~One nontrivial issue, which we shall address in detail, can arise if the state of the system is changing in time between events.~In multiscale (for example within-host and population-level coupled) models, population-level propensities can depend on within-host dynamics, and thus change with time \textit{between} population-level state changes.~The obvious solution here is to attempt to adapt the within-host model into the SSA and generate putative times for all events (including those within each infectious individual).~In this respect, all events, irrespective of whether they are population-level or within-host level would be modelled exactly.~However, it is completely infeasible to do this since:~1) there may be a large number of infectious individuals;~2) each individual will have their own SSA;~and 3) there may be an extremely large number of within-host events simulated before a single population-level model event (the multiscale problem).~Instead, because the within-host dynamics are associated with much smaller timescales compared to the population-level dynamics, it becomes crucial to develop an appropriate method for running an SSA at the population level subject to deterministic and smooth time-varying propensities (in our case as a result of changing viral load of each individual).\\

We are therefore faced with the following two challenges: 1) time varying propensities: propensities that may change with time between discrete event times;~and 2) the individuality problem: temporal change in propensities may be complex due to individual contributions and subject to historical events (for example the previous times when individuals become infected).~The time varying propensity problem has known solutions and can be solved exactly as we shall demonstrate.~Of course, sometimes the additional work required to deal with this problem exactly is not considered to be worth the effort and approximate algorithms, such as $\tau$-leaping, are used (appropriate when the relative change in propensity between events is sufficiently small and therefore approximated to be constant) \cite{gillespie2001approximate}.~The individuality problem is a much harder problem.~This is unsurprising since more detail in SSA simulations and, indeed, any model, will always increase computational demands.~To overcome these two challenges, we introduce in the next section a simple within-host model to use as a test problem and we assume that, once infected, all individuals deterministically trace solutions of this model.~This means that what separates individuals is the initial time of infection only.

\subsubsection{Within-host model}
\label{chap:WH}
In general, a within-host model describes viral dynamics and immune responses at the cellular level in one single host \cite{feng2012model, murillo2013towards}. By assuming that the within-host dynamics can be controlled by target cell limitation only, we choose to rely on an extension of the deterministic target cell-limited model as shown in (\ref{WH ODE}):
\begin{align}
\label{WH ODE}
    \begin{split}
        \frac{dC}{d(\delta t)} &= \Lambda_c - kCV - \mu_cC, \quad C(0) = C_0\\
        \frac{dC^*}{d(\delta t)} &= kCV - (\mu_c + \delta_c)C^*, \quad C^*(0) = C^*_0 \\
        \frac{dV}{d(\delta t)} &= pC^* - cV, \quad V(0) = V_0.
    \end{split}
\end{align}
The independent variable $\delta t$ in the context of the within-host model denotes the age of infection of this infectious host (i.e.~the time since their initial infection). One should keep in mind that $\delta t$ is different from $t$, the population-level calendar time.~The numbers of uninfected cells, infected cells, and virus particles are denoted by $C$, $C^*$, and $V$, respectively.~We assume that the rate at which new target cells are created is given by $\Lambda_c$. The virus is produced at a rate of $p$ while $c$ is its clearance rate, and the infection coefficient of healthy cells is given by $k$.~The constants $\mu_c$ and $\delta_c$ denote, respectively, the mortality rate of the cells and the additional death rate for infectious cells.~The schematic diagram for this within-host model is shown in \autoref{WH_Schematic}.

\subsubsection{Model coupling}
\label{model coupling}
With the specific population-level and within-host sub-models discussed in Section \ref{population level model} and Section \ref{chap:WH}, respectively, we can now construct a multiscale model by linking these separate sub-systems. Theoretically, this coupling process can be facilitated through identifying possible parameters and variables in one sub-system that affect the dynamics of the other and consistently formulating a feedback across sub-models.~A common approach to accomplish this is by expressing that set of parameters or variables in one sub-model as functions of those in the other \cite{garira2014mathematical}. In this section, we will develop a test multiscale model that links the population-level transmission coefficient to the within-host viral load.~It is important to note that, to illustrate the simulation methodology, the model coupling presented below represents just one method to connect these scales.\\

Although a variety of experimental observations \cite{edenborough2012mouse, blaser2014impact} indicate that population-level disease transmission is often associated with the infectious agent's within-host pathogen load, the specific functional form between these quantities is still uncertain \cite{handel2015crossing}.~For simplicity, we assume that each agent’s transmission coefficient is a linear function of the viral load associated with that agent, with proportional constant $l$, which takes the unit $\rm day^{-1}\text{ }viral\, particle^{-1}\text{ }\rm person^{-1}$.~Whilst we have chosen a linear relationship between the viral load $V$ and the agent transmission coefficient, the nature of relationship does not affect at all the applicability of this method.~We recall that $V(\delta t)$ is the viral load of an individual with age of infection $\delta t$.~We shall assume that all infectious individuals within-host dynamics are governed by the same deterministic model (\ref{WH ODE}). Thus, for a specific infectious individual $j$, the corresponding transmission coefficient $\beta_j$ at time $t$ is
\begin{align}
    \label{coupling: beta}
    \beta_j(t) = l V(\delta t_j) = v(t - t_j),
\end{align}
where $v(t - t_j) = lV(\delta t_j)$ and $t - t_j = \delta t_j$. In (\ref{coupling: beta}), $t_j$ is the calendar time associated with the initial infection of individual $j$ and $\delta t_j$ is the age of the infection in that individual.~Note that the individuality of $\beta_j$ is encapsulated only in the initial infection time $t_j$.~We further assume that within-host dynamics are independent of the population-level dynamics (and can be solved numerically once at the start of the simulation).~The function $v(\delta t)$ is assumed to be zero prior to infection ($\delta t<0$), so as to account for the agent being susceptible (noninfectious).\\

A more complicated model can include further stochasticity in the parameters of the within-host model (\ref{WH ODE}) among individuals.~In this case, the appropriate change to the population-level model simply requires noting that $v$ is different for each $j$.~However, introducing more individuality into the model substantially complicates the computational requirements.~Therefore, for a level of novel tractability of the problem, we shall look only at the case where the within-host model is deterministic and the dynamics are the same for each individual post-infection.\\

 Simulating the multiscale coupled model involves replacing the transmission term in (\ref{BH submodel}) with a non-Markovian (explicitly time-dependent) one, which is derived from the sum of the total viral load in the infectious population:
\begin{align}
    \label{transmission term}
    \alpha^{(i)}(s,i) = \beta si \rightarrow \alpha^{(i)}(s,i;t) = s \sum_{j=1}^i \beta_j(t) =  s \sum_{j=1}^i v(t - t_j).
\end{align}
Since the propensity $\alpha^{(i)}$ now depends explicitly on time, the previously mentioned algorithm in Section \ref{chap:SSA} needs to be updated.~Furthermore, it is important to note that this function does not have a simple closed-form expression, as the infection times $\{t_j\}$ are stochastic.\\

\begin{figure}[t]
    \centering   
    \begin{subfigure}[b]{0.3\textwidth}
         \centering
         \includegraphics[width=\textwidth]{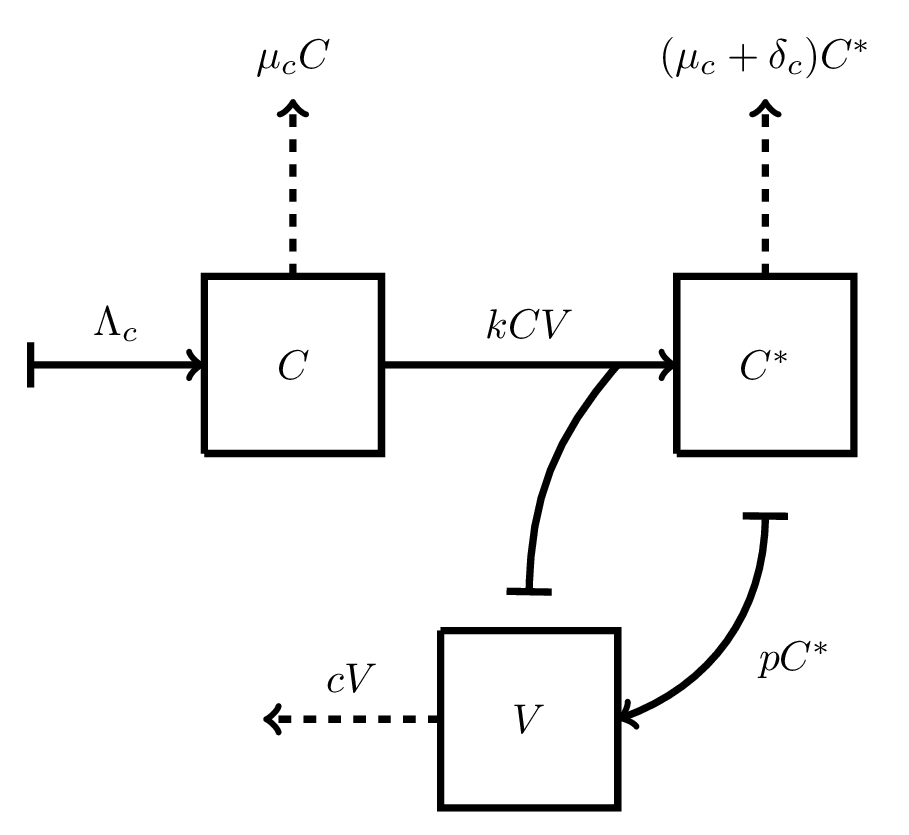}
         \caption{}
         \label{WH_Schematic}
    \end{subfigure}
    \hfill
    \begin{subfigure}[b]{0.22\textwidth}
         \centering
         \includegraphics[width=\textwidth]{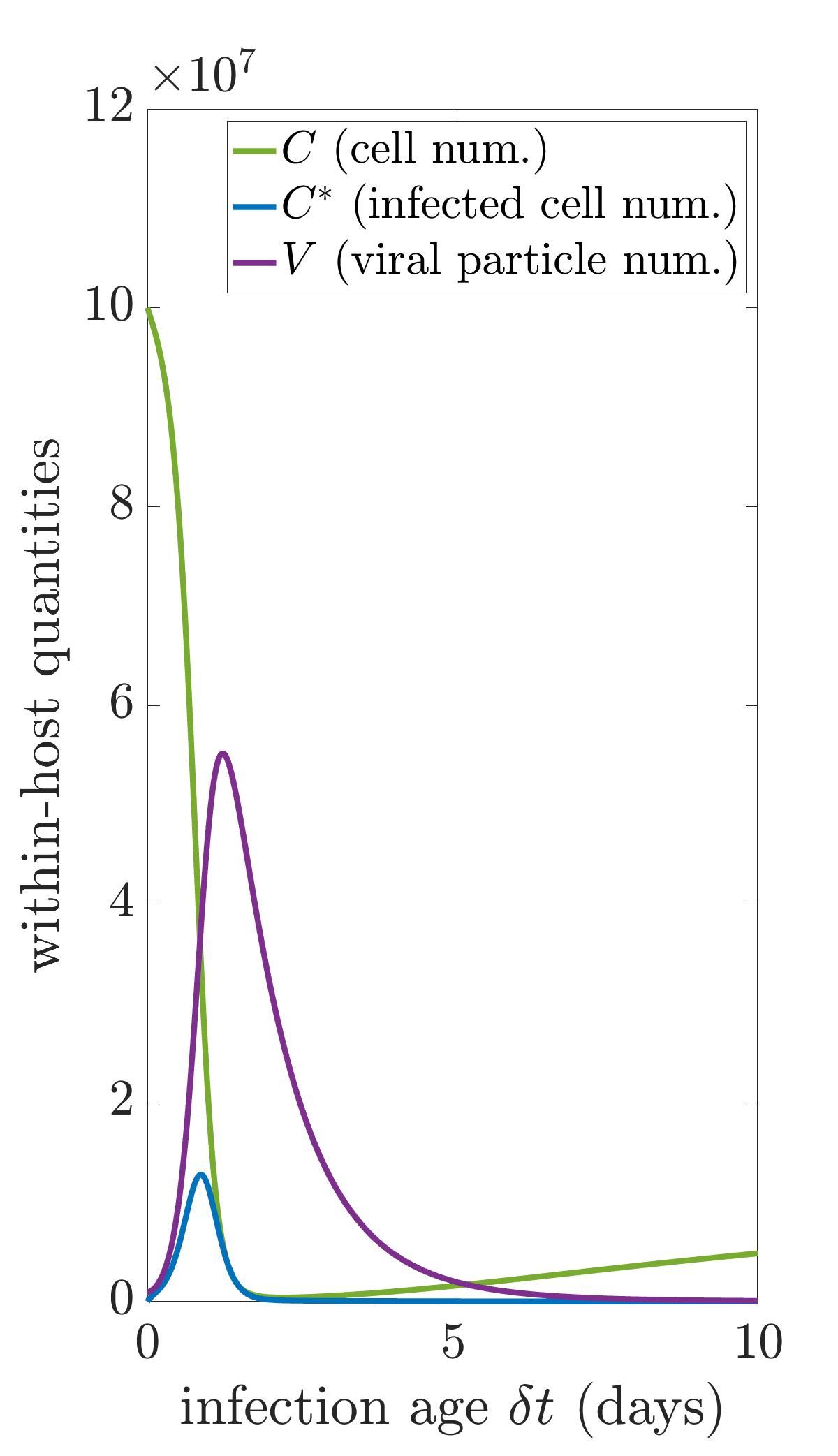}
         \caption{}
         \label{WH dynamics}
    \end{subfigure}
    \hfill
    \begin{subfigure}[b]{0.22\textwidth}
         \centering
         \includegraphics[width=\textwidth]{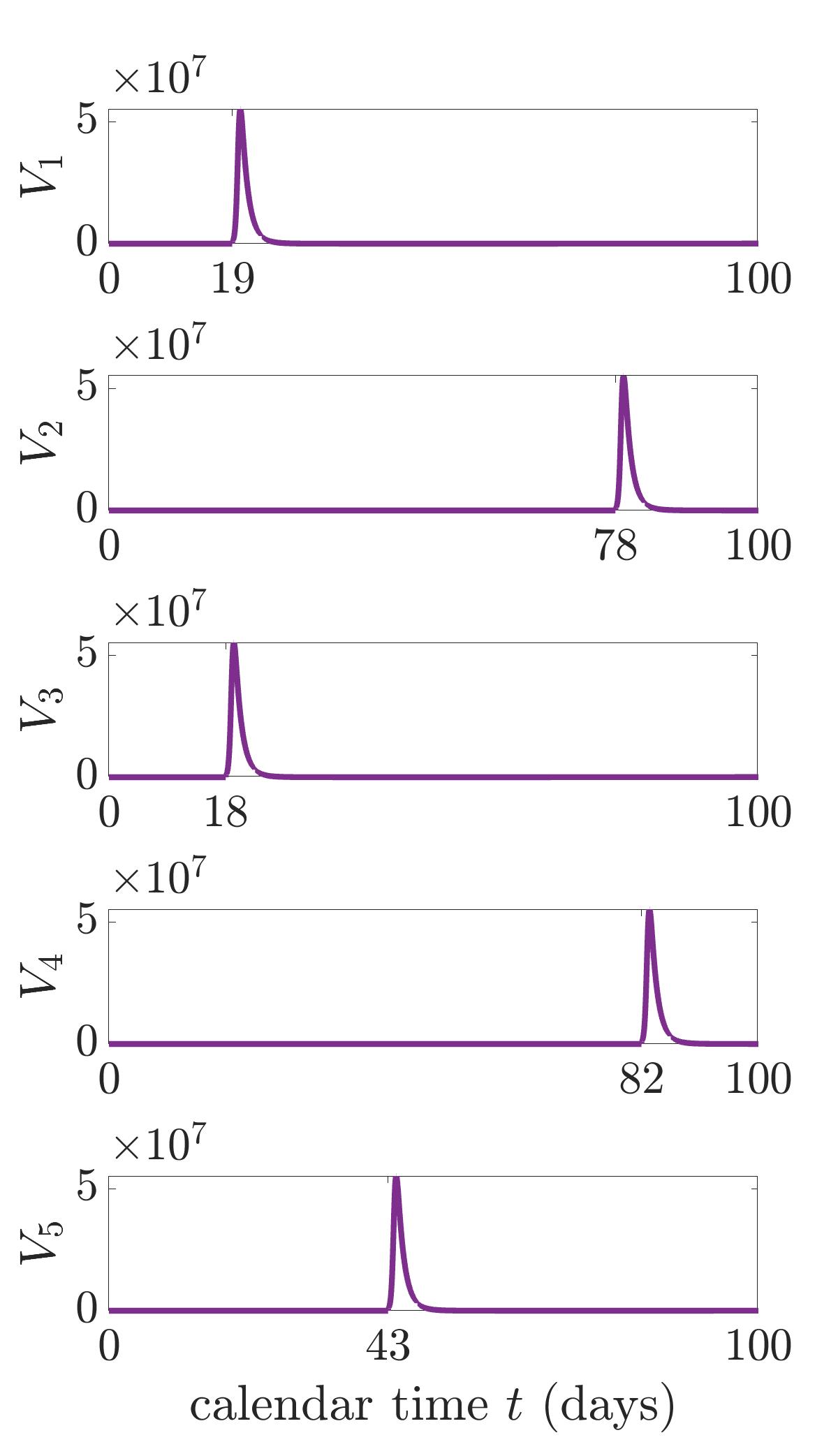}
         \caption{}
         \label{WH 5 dynamics}
    \end{subfigure}
    \begin{subfigure}[b]{0.22\textwidth}
         \centering
         \includegraphics[width=\textwidth]{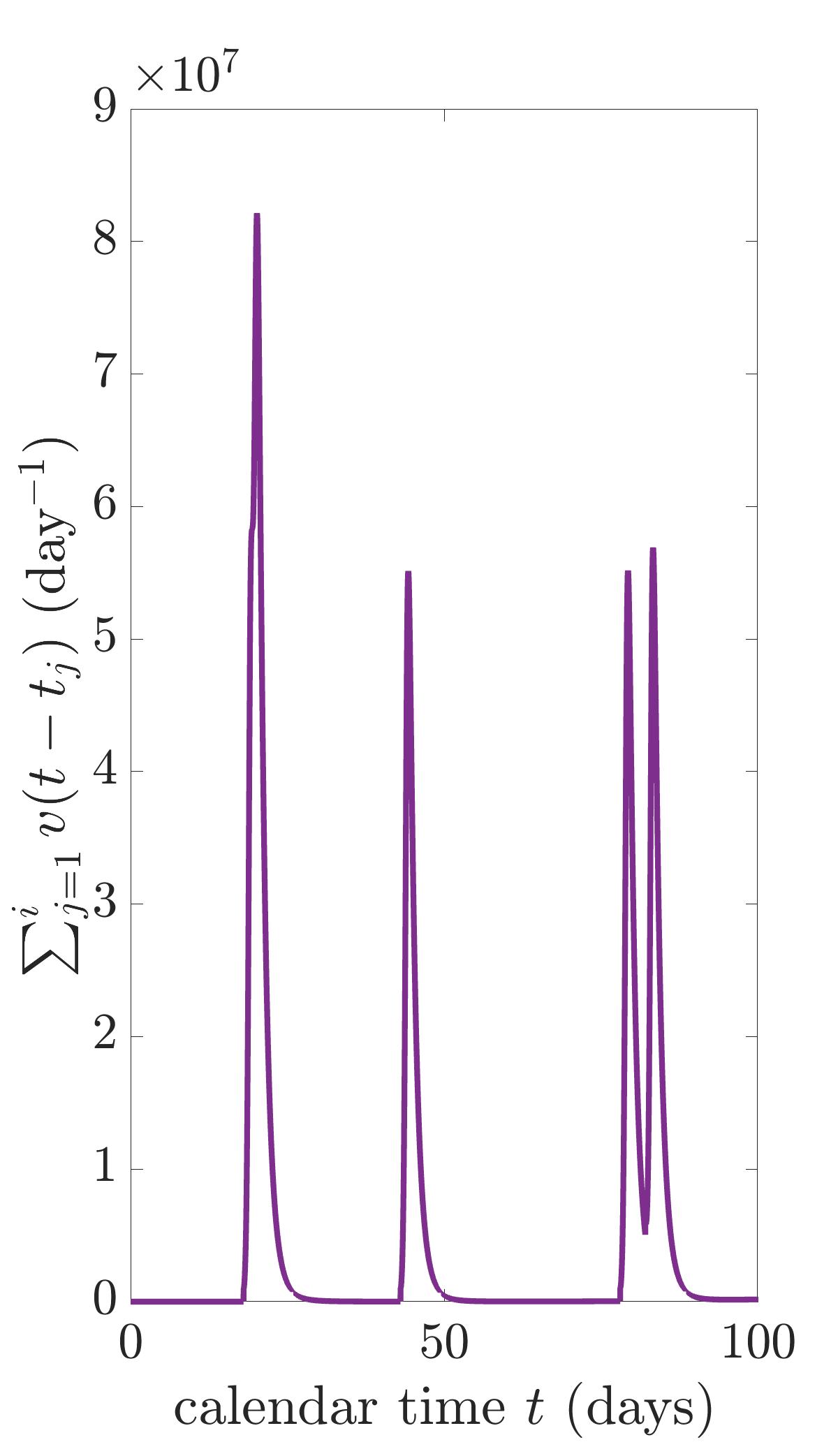}
         \caption{}
         \label{WH overall dynamics}
    \end{subfigure}
    \caption{\textbf{(a) Schematic diagram for the within-host model.} The boxes represent different cell compartments while the arrows denote transitions between compartments.~Note that the compartments linked to the tail end of the arrow, $(\mapsto)$, does not experience any cell nor virus flow in the corresponding transition.~The dashed arrows denote the removal rates.~\textbf{(b) Sample numerical solution of the within-host model (\ref{WH ODE})}.~The solution is plotted as a function of infection age $\delta t$ from MATLAB's $ode15s$ solver with both absolute and relative tolerances being $10^{-8}$.~We set the initial condition $\left(C_0, C_0^*, V_0\right)$ as $(10^8, 10^0, 10^6)$ number of cells or viral particles.~The parameter values chosen are: $p = 10 \rm \text{ }day^{-1}$, $c = 1\rm \text{ }day^{-1}$, $k = 10^{-7}\rm \text{ }day^{-1}\rm cell^{-1}$, $\mu_c = 10^{-1}\rm \text{ }day^{-1}$, $\delta_c = 10\rm \text{ }day^{-1}$\text{, and } $\displaystyle \Lambda_c = (1.111 \times 10^{6}) \text{ cell}\times \text{day}^{-1} $.~\textbf{(c) Viral dynamics for each of 5 individuals}.~These plots also represent the per capita forces of infection ($v(t-t_j)$) divided by the coupling constant ($l$) for each of 5 individuals (see (\ref{coupling: beta})), with different calendar infection dates $\{t_j\} = \{19, 78, 18, 82, 43\}$ respectively generated by the population-level SSA.~They are shown relative to the calendar time $t$.~\textbf{(d) Per susceptible propensity of infection.}~The propensity of an infection per infected individual $\alpha^{(i)}(t)/s$ in the population-level model assumes the complex form of the sum of each individual transmission coefficient by (\ref{transmission term}).~We choose the coupling constant $l$ to take the value of one per day in this plot.}
    \label{WH-system schematic}
\end{figure}

Diagrammatically, the integration of the within-host model is shown in Figure \ref{WH-system schematic}.~The schematic diagram for the within-host model is presented in Figure \ref{WH-system schematic}(a).~In Figure \ref{WH-system schematic}(b) we solve this model to determine the within-host viral load for an individual as a function of its age of infection $\delta t$.~In Figure \ref{WH-system schematic}(c) we use the known viral load from the solution of the within-host model and position the starting time $\delta t = 0$ day to be at the calendar times $t_j$ for each infected individual (of which we plot 5 sample individuals becoming infected at various times) in the population-scale SSA.~Finally, we note that what drives the SSA is the total infectious propensity $\alpha^{(i)}$ as computed by (\ref{transmission term}), the sum of all individual transmission rates in the system multiplied by the susceptible population and presented in Figure \ref{WH-system schematic}(d).

\subsection{Multiscale model simulation}
\label{MultiscaleSSA}
The within-host model (\ref{WH ODE}) can be solved numerically to give $v(\delta t)$.~The multiscale stochastic model can therefore be written in the same manner as (\ref{forward Kolmogorov differential equation for BH submodel}) in Section \ref{chap:SSA}, but noting the change in transmission coefficient as shown in (\ref{transmission term}). Probabilities for updating the simulation from known state $(s,i)$ from (\ref{update eq})-(\ref{update eq4}) in the multiscale case at time $t$ is conditional on the set of times in the past individuals firstly became infectious. That is, by denoting $P_{(s, i)}(t)$ as the probability of being in state $(s,i)$ conditional on known infection times $\{t_j\}_{j = 1}^{i}$:

 \begin{align}
    \label{update eq5} P_{(s,i)}(t + \Delta t) &= 1 - \left(\alpha^{(b)}+\alpha^{(ds)}(s) +\alpha^{(di)}(i)+ s\sum_{j=1}^i v(t-t_j) \right) \Delta t = 1 - \alpha(s,i;t) \Delta t \\
     \label{update eq6}
     P_{(s+1,i)}(t + \Delta t) &= \alpha^{(b)} \Delta t \\
     P_{(s-1,i)}(t + \Delta t) &= \alpha^{(ds)}(s) \Delta t \\
     P_{(s,i-1)}(t + \Delta t) &= \alpha^{(di)}(i) \Delta t \\
      \label{update eq9}
     P_{(s-1,i+1)}(t + \Delta t) &= \alpha^{(i)}(s,i;t)\Delta t = s \sum_{j=1}^i v(t-t_j) \Delta t. 
 \end{align}\\
Applying the strategy used in NRM in the case of the time dependent total propensity $\alpha(s,i;t)$, putative times for birth and deaths can still be found in the standard way as their Markovian property still holds.~However, drawing a putative time for the next infection event requires sampling from a distribution with a time dependent rate.~The derivation of such putative time is to sample the time for a particular event conditional on the state remaining unchanged.~According to (\ref{update eq9}), the probability of an infection in time interval $(t,t+\Delta t)$ is $s\Delta t \sum_{j=1}^i v(t-t_j) + O(\Delta t^2)$. Let $\tilde{p}(n)$ be the probability that no infection occurs in the first $n$ time steps from $t$. We have
$$
\tilde{p}(n) = \prod_{k = 0}^{n-1} \left(1 - \Delta t \gamma(t+k\Delta t) \right) +O(\Delta t^2),
$$
where $\gamma(t) = s \sum_{j=1}^i v(t-t_j)$.
We notice that 
\begin{equation} \label{disctime} \frac{\tilde{p}(n+1) - \tilde{p}(n)}{\Delta t} = -\tilde{p}(n) \gamma(t+n\Delta t). 
\end{equation}

We define the survival distribution function $\text{SDF}_s(T^{(i)}; t)$ as the probability that no infections have occurred in time interval $(t,t+T^{(i)})$ in the limit $\Delta t\rightarrow 0$. By setting $n\Delta t = T^{(i)}$ and taking the limit $\Delta t\rightarrow 0$, we find that $$\frac{\mathrm{d}(\text{SDF}_s(T^{(i)}; t))}{\mathrm{d}T^{(i)}} = -\gamma(t+T^{(i)}) \text{SDF}_s(T^{(i)}; t).$$
Therefore, the survival distribution function as well as its complement, the cumulative distribution function (CDF), associated with the putative time for infection $\tau^{(i)}$ are

\begin{equation}
    \text{SDF}_{s}\left(T^{(i)}; t\right) = 1-\text{CDF}_{s}\left(T^{(i)}; t\right) =  \exp\left[- s \sum_{j=1}^i \Psi(T^{(i)} ; t-t_j) \right].\label{CDF S(t) j2}
\end{equation}
where 
\begin{align}
\label{nu}
    \Psi(T^{(i)};\delta t_j) = \int_{0}^{T^{(i)}} v\left(\eta + \delta t_j \right)d\eta = \Psi(T^{(i)}+\delta t_j)-\Psi(\delta t_j).
\end{align}
Note for notational simplicity, $\Psi(\delta t) \equiv \Psi(\delta t;0)$. By storing $\Psi(\delta t)$ numerically in a look-up table rather than $v(\delta t)$, it is therefore possible to quickly obtain $\Psi(T^{(i)};\delta t_j)$.\\

A sample of the putative time for an infection $\tau^{(i)}$ can be generated by sampling a random number $u\sim U(0,1)$ and solving $\text{CDF}_{s}\left(\tau^{(i)}; t\right) = u$ for $\tau^{(i)}$, where $t$ is the current calendar time.~This is complicated because the exponent in $\text{CDF}_{s}\left(\tau^{(i)}; t\right)$ as shown in (\ref{CDF S(t) j2}) is a nontrivial function which depends on the stochastic history of infections $\{t_j\}_{j=1}^i$ among the infectious population.~Furthermore, even if efficient sampling from the CDF in (\ref{CDF S(t) j2}) is solved, $\tau^{(i)}$ is only a putative time for infection at current calendar time $t$; once the within-host as well as the population-level states change when time forwards and when any event occurs, a new putative time would need to be sampled.~In Section \ref{Naive}, we will firstly outline an approximate computational algorithm that uses discrete time steps $\Delta t$ and equations (\ref{update eq5})-(\ref{update eq9}) to simulate the multiscale scenario.~Since the algorithm propagates forward in time using fixed time steps, they are usually called `time-driven' algorithms.~This is in contrast to the NRM approach which propagates forward in time by the occurrence of events; aptly named `event-driven' approaches.~We will then outline our main result in Section \ref{clever}, which uses a similar strategy as in NRM method where $\tau^{(i)}$ is appropriately resampled and rescaled using the CDF presented in (\ref{CDF S(t) j2}).

\subsubsection{Approximate time-driven simulation}
\label{Naive}

The time-driven algorithm we will present in this section is rather simple but is only approximate.~It uses a predetermined fixed small time step $\Delta t$ with brute force to simulate events using (\ref{update eq5})-(\ref{update eq9}).~Reducing the size of $\Delta t$ allows for control over the level of accuracy but at the cost of efficiency.~The algorithm is a simple extension of the time-driven algorithm previously discussed in Section \ref{chap:SSA}.~We shall use this algorithm to compare with the main algorithm of this paper in Section \ref{clever}.~The pseudocode for this algorithm is shown in Algorithm \ref{Approximate time-driven algorithm}, whereas the MATLAB implementation and all data generated from the simulations are available on Github at \url{https://github.com/YuanYIN99/Accurate-stochastic-simulation-algorithm-for-multiscale-models-of-infectious-diseases.git}.

\begin{algorithm}
  \caption{Approximate time-driven algorithm (omitting units for readability)}
  \begin{algorithmic}[1]
    \State \textbf{Input}: \\
    \quad Solution of individual post-infection force of infection $v(\delta t) = lV(\delta t)$ calculated from a defined within-host model (for example (\ref{WH ODE}));\\
    \quad Population-level parameters (for example, $\Lambda$ and $\mu$ defined by (\ref{BH submodel}));\\
    \quad Initialisation: $t = 0, \; \left(s(0), i(0)\right)=\left(s_0, i_0\right)$: initial conditions for the population-level model;\\
    \quad  Initialisation: $\left\{t_j\right\}_{j = 1}^{i_0} \leq  0 \; \forall j$: times of initial infection for all individuals infectious at $t=0$; \\
    \quad $t_{\mathrm{end}}$: ending time for the simulation.\\

\State Initialise the fixed time stepping size $\Delta t$ (for accurate results ensure that probabilities in (\ref{update eq6})-(\ref{update eq9}) are always small (less than some tolerance $\epsilon$): $\max\left\{\alpha^{(b)} + \alpha^{(ds)} + \alpha^{(di)} + s\sum_{j = 1}^i v(t-t_j)\right\}\Delta t \leq \epsilon, \, \forall t$, where $v(\delta t)$ is defined in (\ref{coupling: beta})). \\

    \State Tabulate $v(\delta t)$ in a vector $\vec{v}\in\mathbb{R}^{n\times 1}$ s.t. the $k$th element is $v_k = v(\frac{(2k-1)\Delta t}{2})$.~Here, $n$ is the largest integer s.t. $\frac{(2n-1)\Delta t}{2} \leq t_{\mathrm{end}}$ (or earlier if $v$ reaches 0 in finite $\delta t$).~We also store an integer vector $\vec{k}\in \mathbb{Z}^{i+s}$ (one element for each individual in the population).~The value of the $j$th element of $\vec{k}$ is initialised and indexes the current (initial) time since first infection of $j$th individual, $k_j = \left[\frac{(t-t_j)}{\Delta t} + \frac{1}{2}\right]$. The first $i$ elements of $\vec{k}$ correspond to the infectious population and therefore $v_{k_j}> 0$ for $1\leq j \leq i$.~The remaining elements of $\vec{k}$ remain zero until respective susceptible individuals are infected. 

    \State
    \While{$t + \Delta t \leq t_{end}$}
      \State Sample $u_1\sim U(0,1)$.
      \State Compute $\alpha$ based on (\ref{update eq5}), where $\sum_{j=1}^i v(t-t_j) = \sum_{j = 1}^i {v}_{k_j}$ by (\ref{transmission term}).
      \If{$u_1 < \alpha$}
         \State Sample another random number $u_2\sim U(0,1)$ to determine which event occurs according to probabilities (\ref{update eq6})-(\ref{update eq9}). 
         \State Update the population-level state $(s,\, i)$ accordingly.
         \If{the event is death}
         \State Remove a susceptible or infectious person (based on event) randomly from $\vec{k}$.  
         \ElsIf{the event is introduction of one susceptible}
         \State Append a zero on the end of $\vec{k}$.
         \EndIf
      \EndIf
      \State Update $t := t + \Delta t$.
      \State $k_j:= k_j + 1$, for each $j$ corresponding to an infected individual, $ 1 \leq j \leq i$.
    \EndWhile\\
    \State \textbf{Return} $s$ and $i$ at each time step.
  \end{algorithmic}
  \label{Approximate time-driven algorithm}
\end{algorithm}

\subsubsection{Accurate multiscale simulation}
\label{clever}
Based on the strategy in NRM, the event with the minimum putative time is selected and the state $(s,i)$ is updated to $(\bar{s},\bar{i})$ corresponding to the change caused by that event.~After changing the state, the time is first updated and then the putative time for the event that occurred is resampled.~For all other events, the putative time is updated to reflect the time that has passed and then rescaled to reflect any change in propensity as a result of the transition $(s,i)$ to $(\bar{s},\bar{i})$.~In this section, we present an algorithm based on the idea of NRM, for simulating non-Markovian disease transmission models of the type explored in this manuscript.~We focus on how putative times corresponding to within-host infections $\tau^{(i)}$ are computed, resampled, and rescaled to be used in the general next reaction method framework. \\

To find $\tau^{(i)}$ we could attempt to solve $\mathrm{SDF}_s(\tau^{(i)};t) = u$ where $u= 1-\tilde{u} \sim U(0,1) $, where $\mathrm{SDF}_s$ is defined in (\ref{CDF S(t) j2}).~However, instead of finding a putative time $\tau^{(i)}$ for the next infection, we note that infected individuals are unique in this model (at least when it comes to infecting others) and therefore it is more tractable to subdivide infection events by the individual who is responsible for the infection; and find a putative infection time for each infected individual $j$ as $\tau_j^{(i)}$.~We shall then use the minimum of these putative times, based on the idea from NRM, to determine $\tau^{(i)} = \mathrm{min}_j (\tau_j^{(i)})$. In fact, going further than this we determine the update time $\tau^{(m)} = \mathrm{min} (\{\tau_j^{(i)}\}, \tau^{(di)}, \tau^{(ds)},\tau^{(b)})$.~Using this method of treating each individual infected person separately, we also have the capability of storing data on who is responsible for distributing the disease for later analysis without any further calculation. \\

We are permitted to (and indeed benefited by) applying the strategy of NRM in this way because
\begin{equation}
\mathrm{SDF}_s(T^{(i)};t)  =  \mathbb{P}\left[\tau^{(i)} > T^{(i)} \right].
\end{equation}
We denote the survival probability for an infection by infectious individual $j$ to be 
\begin{equation} \label{infectedindividual}
       \text{SDF}_{s;j}\left(T^{(i)}; t\right) = \mathbb{P}\left[\tau_j^{(i)} > T^{(i)} \right] = \exp\left[- s \Psi(T^{(i)} ; t-t_j) \right],
\end{equation}
and similarly $\text{CDF}_{s;j} = 1-\text{SDF}_{s;j}$.~Note that $1 \leq j \leq i$; $t-t_j$ denotes the age of infection; $\Psi$ is defined in (\ref{nu}). Let $\tau_m^{(i)} = \mathrm{min}_j (\tau_j^{(i)})$.~The idea of NRM asserts that $\tau_m^{(i)}$ and $\tau^{(i)}$ have the same distribution, which can easily be seen by
$$
\mathbb{P}\left[\tau_m^{(i)} > T^{(i)} \right] = \mathbb{P} \left[\cap_{j=1}^i (\tau_j^{(i)} > T^{(i)}) \right] = \prod_{j=1}^i \mathbb{P}\left[\tau_j^{(i)} > T^{(i)} \right] = \mathrm{SDF}_s(T^{(i)};t)=\mathbb{P}\left[\tau^{(i)} > T^{(i)} \right].
$$
Note that because each of the putative times $\tau_j^{(i)}$ are independently sampled, the survival distribution function for the next infection time (\ref{CDF S(t) j2}) is the product of that for each infectious person (\ref{infectedindividual}). Therefore, by breaking down the infection event by infectious individual, we do not have to compile the information about the distribution of viral loads in the population to calculate a putative time for an infection. \\

As in the NRM, when $\tau^{(m)} = \tau_j^{(i)}$, the `next event' is infection by the $j$th infected individual.~For this individual, the next infection will need to be \textit{resampled}.~For all the other putative times, $\tau_k^{(i)}>\tau^{(m)}$ where $k\neq j$, we apply a similar strategy as in NRM for \textit{rescaling} of the putative time instead of resampling.~We describe both resampling and rescaling next in detail. \\

\textbf{Resampling $\tau_j^{(i)}$}

Drawing a putative time from the survival distribution function (\ref{infectedindividual}) for each infected individual is achieved by first sampling $u_j \sim U(0,1)$ and using a look up table to solve for $\tau_j^{(i)}$ where $u_j = \mathrm{SDF}_{s;j}(\tau_j^{(i)};t)$.~To achieve this, it makes sense when looking at (\ref{infectedindividual}) to focus on the variable $\psi_j = -\log(u_j)/s$ instead of $u_j$ since this allows us to tabulate a function that is independent of $s$ as follows. Using (\ref{infectedindividual}) and recalling that $\Psi(\tau_j^{(i)};\delta t_j) = \Psi(\tau_j^{(i)}+\delta t_j) - \Psi(\delta t_j)$,

\begin{equation}
\label{xi}
    \psi_j= - \frac{\log(u_j)}{s}  = \Psi(\tau_j^{(i)} + \delta t_j) - \Psi(\delta t_j).
\end{equation}
That is, the value of $\psi_j = -\log(u_j)/s$ gives the putative increase in the function $\Psi$ (which is independent of $s$) from the current time to the time of the infection.~We therefore track four values for each infected individual: (1) the current value of the individuals infection time $\delta t_j$, (2) the current putative time until the individual next infects a susceptible $\tau_j^{(i)}$, (3) the current value of $\Psi_j = \Psi(\delta t_j)$, and 
 (4) the current putative change the value of $\Psi_j$ when the individual next infects $\psi_j = \Psi(\tau_j^{(i)} + \delta t_j) - \Psi_j(\delta t_j)$.~It is the later putative variable $\psi_j=-\log(u_j)/s$ (not the putative time $\tau_j^{(i)}$) which can easily be sampled.~However, as time moves forward and infection time needs updating, this has to be done in the time domain by shifting $\delta t_j$.
Resampling should be done if $\tau^{(m)}=\tau_j^{(i)}$. \\

In the case of resampling, time should be first updated to the new infection time 
 \begin{equation}\label{timeupdate}
     \bar{\delta t_j} = \delta t_j + \tau^{(m)},
 \end{equation}
where the bar notation indicates the `new' value of the variable.~The update step also allows for the calculation of the new $\Psi_j$, that is $\bar{\Psi_j} = \Psi(\bar{\delta t_j})$ by the look-up table for the function $\Psi$.~Second, after the change of state $(s,i)$ to $(\bar{s},\bar{i})$ a new putative $\psi$ is sampled for the individual; $\bar{\psi_j} = -\log{u_j}/\bar{s}$ by (\ref{xi}) reflecting the new $\bar{s}$. Finally, it is possible to determine the new putative time $\bar{\tau_j^{(i)}}$. This is done simply by knowing $\bar{\delta t_j}$, $\bar{\Psi_j}$ and $\bar{\psi_j}$ and using the look-up table to invert the function $\Psi$:
\begin{equation}\label{tauupdate}
\bar{\tau_j^{(i)}} = \Psi^{-1}(\bar{\Psi_j}+\bar{\psi_j}) - \bar{\delta t_j}.
\end{equation}
\\

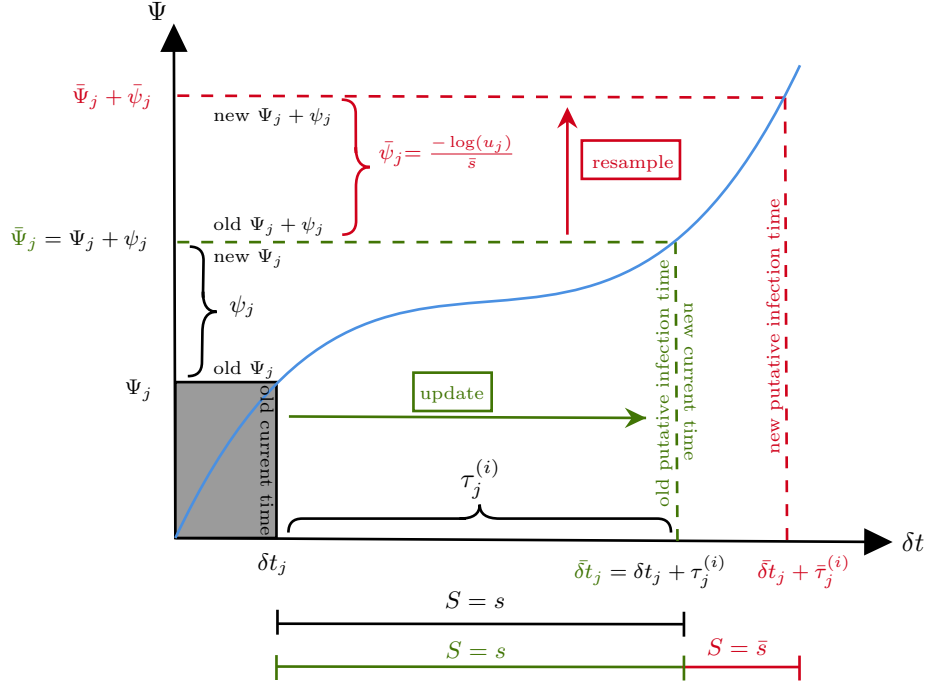
\begin{figure}[h!]
\centering
\tikzset{every picture/.style={line width=1pt}} 
\begin{tikzpicture}[x=0.75pt,y=0.75pt,yscale=-1,xscale=1]
\draw    (121,276) -- (478,277.98) ;
\draw [shift={(481,278)}, rotate = 180.32] [fill={rgb, 255:red, 0; green, 0; blue, 0 }  ][line width=0.08]  [draw opacity=0] (14.29,-6.86) -- (0,0) -- (14.29,6.86) -- cycle    ;
\draw    (121,276) -- (120.26,20.75) ;
\draw [shift={(120.25,17.75)}, rotate = 89.83] [fill={rgb, 255:red, 0; green, 0; blue, 0 }  ][line width=0.08]  [draw opacity=0] (14.29,-6.86) -- (0,0) -- (14.29,6.86) -- cycle    ;
\draw [color={rgb, 255:red, 65; green, 117; blue, 5 }  ,draw opacity=1 ] [dash pattern={on 4.5pt off 4.5pt}]  (122,127.5) -- (372,127.5) ;
\draw [color={rgb, 255:red, 65; green, 117; blue, 5 }  ,draw opacity=1 ][fill={rgb, 255:red, 65; green, 117; blue, 5 }  ,fill opacity=1 ] [dash pattern={on 4.5pt off 4.5pt}]  (372,127.5) -- (373,278) ;
\draw [color={rgb, 255:red, 208; green, 2; blue, 27 }  ,draw opacity=1 ] [dash pattern={on 4.5pt off 4.5pt}]  (122,54) -- (427,55) ;
\draw [color={rgb, 255:red, 208; green, 2; blue, 27 }  ,draw opacity=1 ] [dash pattern={on 4.5pt off 4.5pt}]  (427,55) -- (428,279) ;
\draw   (127.25,194.75) .. controls (131.92,194.82) and (134.29,192.53) .. (134.36,187.86) -- (134.59,172.36) .. controls (134.7,165.69) and (137.08,162.4) .. (141.75,162.47) .. controls (137.08,162.4) and (134.8,159.03) .. (134.9,152.36)(134.86,155.36) -- (135.14,136.86) .. controls (135.21,132.19) and (132.92,129.82) .. (128.25,129.75) ;
\draw  [color={rgb, 255:red, 208; green, 2; blue, 27 }  ,draw opacity=1 ] (204.75,123.75) .. controls (209.42,123.73) and (211.74,121.39) .. (211.72,116.72) -- (211.66,99.85) .. controls (211.64,93.18) and (213.96,89.84) .. (218.63,89.82) .. controls (213.96,89.84) and (211.62,86.52) .. (211.59,79.85)(211.6,82.85) -- (211.53,62.97) .. controls (211.51,58.3) and (209.17,55.98) .. (204.5,56) ;
\draw   (369.75,273.25) .. controls (369.75,268.58) and (367.42,266.25) .. (362.75,266.25) -- (283.75,266.25) .. controls (277.08,266.25) and (273.75,263.92) .. (273.75,259.25) .. controls (273.75,263.92) and (270.42,266.25) .. (263.75,266.25)(266.75,266.25) -- (184.75,266.25) .. controls (180.08,266.25) and (177.75,268.58) .. (177.75,273.25) ;
\draw  [fill={rgb, 255:red, 155; green, 155; blue, 155 }  ,fill opacity=1 ] (121,197.75) -- (171.75,197.75) -- (171.75,276) -- (121,276) -- cycle ;
\draw    (172.25,319) -- (375.75,319.25) ;
\draw [color={rgb, 255:red, 65; green, 117; blue, 5 }  ,draw opacity=1 ]   (171.75,340.5) -- (375.25,340.75) ;
\draw [color={rgb, 255:red, 208; green, 2; blue, 27 }  ,draw opacity=1 ]   (375.25,340.75) -- (434.25,340.75) ;
\draw [color={rgb, 255:red, 65; green, 117; blue, 5 }  ,draw opacity=1 ]   (171.75,334.75) -- (171.75,347.25) ;
\draw    (172.25,312) -- (172.25,324.5) ;
\draw    (376.25,313) -- (376.25,325.5) ;
\draw [color={rgb, 255:red, 65; green, 117; blue, 5 }  ,draw opacity=1 ]   (376.25,335) -- (376.25,347.5) ;
\draw [color={rgb, 255:red, 208; green, 2; blue, 27 }  ,draw opacity=1 ]   (434.25,335) -- (434.25,347.5) ;
\draw [color={rgb, 255:red, 65; green, 117; blue, 5 }  ,draw opacity=1 ]   (178,215) -- (357.25,215.74) ;
\draw [shift={(357.25,215.75)}, rotate = 180.24] [fill={rgb, 255:red, 65; green, 117; blue, 5 }  ,fill opacity=1 ][line width=0.08]  [draw opacity=0] (10.72,-5.15) -- (0,0) -- (10.72,5.15) -- (7.12,0) -- cycle    ;
\draw  [color={rgb, 255:red, 65; green, 117; blue, 5 }  ,draw opacity=1 ] (240.5,193) -- (279.75,193) -- (279.75,212.25) -- (240.5,212.25) -- cycle ;
\draw [color={rgb, 255:red, 208; green, 2; blue, 27 }  ,draw opacity=1 ]   (317.75,123.75) -- (317.75,62.75) ;
\draw [shift={(317.75,59.75)}, rotate = 90] [fill={rgb, 255:red, 208; green, 2; blue, 27 }  ,fill opacity=1 ][line width=0.08]  [draw opacity=0] (10.72,-5.15) -- (0,0) -- (10.72,5.15) -- (7.12,0) -- cycle    ;
\draw  [color={rgb, 255:red, 208; green, 2; blue, 27 }  ,draw opacity=1 ] (325,76.5) -- (374.25,76.5) -- (374.25,95.75) -- (325,95.75) -- cycle ;
\draw  [color={rgb, 255:red, 74; green, 144; blue, 226 }  ,draw opacity=1 ] (121,276) .. controls (225.5,55.36) and (330,259.39) .. (434.5,38.75) ;

\draw (160.5,281.9) node [anchor=north west][inner sep=0.75pt]  [font=\small]  {$\delta t_{j}$};
\draw (484,271.4) node [anchor=north west][inner sep=0.75pt]    {$\delta t$};
\draw (105,5.4) node [anchor=north west][inner sep=0.75pt]    {$\Psi $};
\draw (94,194.9) node [anchor=north west][inner sep=0.75pt]  [font=\footnotesize]  {$\Psi_{j} \ $};
\draw (36.5,118.4) node [anchor=north west][inner sep=0.75pt]  [font=\footnotesize]  {$\textcolor[rgb]{0.25,0.46,0.02}{\bar{\Psi}_{j}} =\Psi_{j} +\psi_{j} \ $};
\draw (67.5,46.9) node [anchor=north west][inner sep=0.75pt]  [font=\footnotesize]  {$\textcolor[rgb]{0.82,0.01,0.11}{\bar{\Psi}_{j} +\bar{\psi}_{j} \ }$};
\draw (262.5,234.9) node [anchor=north west][inner sep=0.75pt]    {$\tau_{j}^{(i)}$};
\draw (146.5,152.4) node [anchor=north west][inner sep=0.75pt]  [font=\small]  {$\psi_{j} \ $};
\draw (360.5,262) node [anchor=north west][inner sep=0.75pt]  [color={rgb, 255:red, 65; green, 117; blue, 5 }  ,opacity=1 ,rotate=-270] [align=left] {{\scriptsize old putative infection time}};
\draw (384.5,156) node [anchor=north west][inner sep=0.75pt]  [color={rgb, 255:red, 65; green, 117; blue, 5 }  ,opacity=1 ,rotate=-90] [align=left] {{\scriptsize new current time}};
\draw (416,235.5) node [anchor=north west][inner sep=0.75pt]  [font=\scriptsize,color={rgb, 255:red, 208; green, 2; blue, 27 }  ,opacity=1 ,rotate=-270] [align=left] {new putative infection time\\};
\draw (319.5,282.9) node [anchor=north west][inner sep=0.75pt]  [font=\footnotesize]  {$\textcolor[rgb]{0.25,0.46,0.02}{\bar{\delta t}_{j}} =\delta t_{j} +\tau_{j}^{( i)}$};
\draw (172,197.5) node [anchor=north west][inner sep=0.75pt]  [color={rgb, 255:red, 0; green, 0; blue, 0 }  ,opacity=1 ,rotate=-90] [align=left] {{\scriptsize old current time}};
\draw (411.5,282.9) node [anchor=north west][inner sep=0.75pt]  [font=\footnotesize]  {$\textcolor[rgb]{0.82,0.01,0.11}{\bar{\delta t}_{j} +\bar{\tau}_{j}^{( i)}}$};
\draw (254.5,301.4) node [anchor=north west][inner sep=0.75pt]  [font=\small]  {$S=s$};
\draw (255,326.4) node [anchor=north west][inner sep=0.75pt]  [font=\small,color={rgb, 255:red, 65; green, 117; blue, 5 }  ,opacity=1 ]  {$S=s$};
\draw (386,324.9) node [anchor=north west][inner sep=0.75pt]  [font=\small,color={rgb, 255:red, 208; green, 2; blue, 27 }  ,opacity=1 ]  {$S=\bar{s}$};
\draw (242,198) node [anchor=north west][inner sep=0.75pt]  [font=\scriptsize,color={rgb, 255:red, 65; green, 117; blue, 5 }  ,opacity=1 ] [align=left] {update};
\draw (328.5,82) node [anchor=north west][inner sep=0.75pt]  [font=\scriptsize,color={rgb, 255:red, 208; green, 2; blue, 27 }  ,opacity=1 ] [align=left] {resample};
\draw (221.5,71.9) node [anchor=north west][inner sep=0.75pt]  [font=\footnotesize]  {$\textcolor[rgb]{0.82,0.01,0.11}{\bar{\psi}_{j}}\textcolor[rgb]{0.82,0.01,0.11}{=\frac{\textcolor[rgb]{0.82,0.01,0.11}{-\log}\textcolor[rgb]{0.82,0.01,0.11}{(}\textcolor[rgb]{0.82,0.01,0.11}{u}\textcolor[rgb]{0.82,0.01,0.11}{_{j}}\textcolor[rgb]{0.82,0.01,0.11}{)}}{\bar{s}} \ }$};
\draw (138.5,129.4) node [anchor=north west][inner sep=0.75pt]  [font=\scriptsize]  {new $\Psi_{j}$};
\draw (138.5,185.4) node [anchor=north west][inner sep=0.75pt]  [font=\scriptsize]  {old $\Psi_{j}$};
\draw (138.5,113.4) node [anchor=north west][inner sep=0.75pt]  [font=\scriptsize]  {old $\Psi_{j} +\psi_{j}$};
\draw (138.5,58.4) node [anchor=north west][inner sep=0.75pt]  [font=\scriptsize]  {new $ \Psi_{j} +\psi_{j}$};
\end{tikzpicture}
\caption{\textbf{Schematic describing the process of putative infection time resampling for a given individual $j$.}~The blue curve corresponds to the tabulated function $\Psi(\delta t)$ described by (\ref{nu}) and is fixed for all individuals.~The green components indicate updating to the current time within the infection process for this individual and the red components indicate the resampling process for the next infection time.~All characters without bars indicate original parameter values and overbars indicate that they are updated values.}\label{resamplingfigure}
\end{figure}

A schematic showing how $\tau^{(i)}_j$ is resampled is presented in Figure \ref{resamplingfigure}.~Here we show how the update in time (green) is to the putative time for the individual $j$ ($\tau^{(m)} = \tau^{(i)}_j$) and the time is first updated.~By the green dashed line, it can be seen how the new infection time for the $j$th individual, $\bar{\delta t}_j$, corresponds to the new value $\bar{\Psi}_j$ for this infectious individual.~From here, the resample step is given in red.~A putative future increment in $\Psi$ ($\bar{\psi_j} = -\log(u_j)/\bar{s}$) is computed and then (by the red dashed line) a new putative time until next infection $\bar{\tau}_j^{(i)}$ is determined.  

\newpage
\textbf{Rescaling $\tau_j^{(i)}$}

If $\tau^{(m)}<\tau_j^{(i)}$ naively, it is possible to simply update $\delta t_j$, $\tau_j^{(i)}$, $\Psi_j$ and $\psi_j$ using the resampling method.~Whilst random number generation is cheap in comparison to looking up the value of a nonlinear function and its inverse, it is possible to determine the new putative time $\tau_j^{(i)}$ without generating any new random numbers and only at the cost of a few extra floating point operations.~This can be achieved by simply rescaling $\psi_j$.~We begin by determining, $\bar{\delta t_j}$ and $\bar{\Psi_j}$ as previously described.~We note however that since $\tau_j^{(i)}>\tau^{(m)}$, there is a portion of time between $\bar{\delta t_j}$ and $\delta t_j + \tau_j^{(i)}$ for which the propensity is changed due to the change of state $\bar{s}$ from its original value $s$.~This corresponds to the interval in $\Psi$ of width
\begin{equation}
\psi_j' = \Psi_j + \psi_j - \bar{\Psi_j}>0.
\end{equation}
It is possible to reuse and rescale $\psi_j'$ to generate $\bar{\psi_j}$ such that it has the same probability density as it would if it were resampled using $-\log(u_j)/\bar{s}$.~This is achieved by using the rescaling
\begin{equation}\label{rescalingpsi}
    \bar{\psi}_j = \frac{s}{\bar{s}} \psi_j'.
\end{equation}
This reuse of the random number $\psi'_j$ is possible since

\begin{equation}
   \mathbb{P}\left[\tau_j^{(i)} > T^{(i)} | \tau_j^{(i)} > \tau^{(m)} \right] = \frac{\mathrm{SDF}_{s;j}(T^{(i)};t)}{\mathrm{SDF}_{s;j}(\tau^{(m)};t)} = \mathrm{SDF}_{s;j}(T^{(i)'};t')  = \mathbb{P}\left[\tau_j^{(i)'} > T^{(i)'} \right],
\end{equation}
where $T^{(i)'} = T^{(i)}-\tau^{(m)}$ and $\tau_j^{(i)'} = \tau_j^{(i)}-\tau^{(m)}$ is the wait time from $\bar{\delta t_j}$ instead of $\delta t_j$, indicating an update in the calendar time.~Of course, the probability of finding the updated putative time $\tau_j^{(i)'}$ assumes that the value of $s$ is not changed to $\bar{s}$ at the time $\bar{\delta t_j}$.~We therefore note that $\psi_j'$ can be generated with its correct probability by $-\log(u_j)/s$.~The true updated $\bar{\psi_j}$ should instead use the correct (new) susceptible population $\bar{s}$.~That is, $\bar{\psi}_j = -\log(u_j)/\bar{s} = s\psi_j'/\bar{s}$.~The value of the updated putative time is then calculated in the standard way using (\ref{tauupdate}).\\

A schematic showing how $\tau^{(i)}_j$ is updated and then rescaled after each event (that takes place at the infection time of $\delta t_j + \tau_j^{(m)}$ in the schematic) is presented in Figure \ref{rescaleImage}.~Here we show how the update in time to the next putative time $\tau^{(m)}$ (green).~The new current infection time for this individual $\bar{\delta t}_j$ is then related to the new parameter $\bar{\Psi}_j$ for this individual.~The remaining parameter $\psi_j'$ separating the new $\bar{\Psi}_j$ and the old putative parameter $\Psi_j + \psi_j$ is indicated.~In red, this putative parameter $\psi_j'$ is then scaled to give a new putative value of $\bar{\psi}_j$ which in turn is used to generate the putative time $\bar{\tau_j^{(i)}}$ using a look-up table.

\begin{figure}[t]
\centering
\tikzset{every picture/.style={line width=1pt}} 
\begin{tikzpicture}[x=0.75pt,y=0.75pt,yscale=-1,xscale=1]
\draw    (170,296) -- (527,297.98) ;
\draw [shift={(530,298)}, rotate = 180.32] [fill={rgb, 255:red, 0; green, 0; blue, 0 }  ][line width=0.08]  [draw opacity=0] (14.29,-6.86) -- (0,0) -- (14.29,6.86) -- cycle    ;
\draw    (170,296) -- (169.26,40.75) ;
\draw [shift={(169.25,37.75)}, rotate = 89.83] [fill={rgb, 255:red, 0; green, 0; blue, 0 }  ][line width=0.08]  [draw opacity=0] (14.29,-6.86) -- (0,0) -- (14.29,6.86) -- cycle    ;
\draw [color={rgb, 255:red, 0; green, 0; blue, 0 }  ,draw opacity=1 ] [dash pattern={on 4.5pt off 4.5pt}]  (171,147.5) -- (421,147.5) ;
\draw [color={rgb, 255:red, 0; green, 0; blue, 0 }  ,draw opacity=1 ][fill={rgb, 255:red, 65; green, 117; blue, 5 }  ,fill opacity=1 ] [dash pattern={on 4.5pt off 4.5pt}]  (421,147.5) -- (422,298) ;
\draw [color={rgb, 255:red, 208; green, 2; blue, 27 }  ,draw opacity=1 ] [dash pattern={on 4.5pt off 4.5pt}]  (171,74) -- (476,75) ;
\draw [color={rgb, 255:red, 208; green, 2; blue, 27 }  ,draw opacity=1 ] [dash pattern={on 4.5pt off 4.5pt}]  (476,75) -- (477,299) ;
\draw   (173.75,214.75) .. controls (178.42,214.82) and (180.79,212.53) .. (180.86,207.86) -- (181.09,192.36) .. controls (181.2,185.69) and (183.58,182.4) .. (188.25,182.47) .. controls (183.58,182.4) and (181.3,179.03) .. (181.4,172.36)(181.36,175.36) -- (181.64,156.86) .. controls (181.71,152.19) and (179.42,149.82) .. (174.75,149.75) ;
\draw  [color={rgb, 255:red, 208; green, 2; blue, 27 }  ,draw opacity=1 ] (286.5,172.25) .. controls (291.17,172.25) and (293.5,169.92) .. (293.5,165.25) -- (293.5,134.13) .. controls (293.5,127.46) and (295.83,124.13) .. (300.5,124.13) .. controls (295.83,124.13) and (293.5,120.8) .. (293.5,114.13)(293.5,117.13) -- (293.5,83) .. controls (293.5,78.33) and (291.17,76) .. (286.5,76) ;
\draw   (418.75,293.25) .. controls (418.75,288.58) and (416.42,286.25) .. (411.75,286.25) -- (332.75,286.25) .. controls (326.08,286.25) and (322.75,283.92) .. (322.75,279.25) .. controls (322.75,283.92) and (319.42,286.25) .. (312.75,286.25)(315.75,286.25) -- (233.75,286.25) .. controls (229.08,286.25) and (226.75,288.58) .. (226.75,293.25) ;
\draw  [fill={rgb, 255:red, 155; green, 155; blue, 155 }  ,fill opacity=1 ] (170,217.75) -- (220.75,217.75) -- (220.75,296) -- (170,296) -- cycle ;
\draw    (221.25,339) -- (424.75,339.25) ;
\draw [color={rgb, 255:red, 65; green, 117; blue, 5 }  ,draw opacity=1 ]   (220.75,360.5) -- (336,361.25) ;
\draw [color={rgb, 255:red, 208; green, 2; blue, 27 }  ,draw opacity=1 ]   (336,361.25) -- (483,361.25) ;
\draw [color={rgb, 255:red, 65; green, 117; blue, 5 }  ,draw opacity=1 ]   (220.75,354.75) -- (220.75,367.25) ;
\draw    (221.25,332) -- (221.25,344.5) ;
\draw    (425.25,333) -- (425.25,345.5) ;
\draw [color={rgb, 255:red, 65; green, 117; blue, 5 }  ,draw opacity=1 ]   (336.5,354.75) -- (336.5,367.25) ;
\draw [color={rgb, 255:red, 208; green, 2; blue, 27 }  ,draw opacity=1 ]   (483.25,355) -- (483.25,367.5) ;
\draw [color={rgb, 255:red, 65; green, 117; blue, 5 }  ,draw opacity=1 ]   (227,235) -- (331,234.76) ;
\draw [shift={(334,234.75)}, rotate = 179.87] [fill={rgb, 255:red, 65; green, 117; blue, 5 }  ,fill opacity=1 ][line width=0.08]  [draw opacity=0] (10.72,-5.15) -- (0,0) -- (10.72,5.15) -- (7.12,0) -- cycle    ;
\draw  [color={rgb, 255:red, 65; green, 117; blue, 5 }  ,draw opacity=1 ] (261.5,212) -- (300.75,212) -- (300.75,231.25) -- (261.5,231.25) -- cycle ;
\draw [color={rgb, 255:red, 208; green, 2; blue, 27 }  ,draw opacity=1 ]   (366.75,143.75) -- (366.99,81.25) ;
\draw [shift={(367,78.25)}, rotate = 90.22] [fill={rgb, 255:red, 208; green, 2; blue, 27 }  ,fill opacity=1 ][line width=0.08]  [draw opacity=0] (10.72,-5.15) -- (0,0) -- (10.72,5.15) -- (7.12,0) -- cycle    ;
\draw  [color={rgb, 255:red, 208; green, 2; blue, 27 }  ,draw opacity=1 ] (375.5,96.5) -- (416,96.5) -- (416,115.75) -- (375.5,115.75) -- cycle ;
\draw  [color={rgb, 255:red, 74; green, 144; blue, 226 }  ,draw opacity=1 ] (170,296) .. controls (274.5,75.36) and (379,279.39) .. (483.5,58.75) ;
\draw [color={rgb, 255:red, 65; green, 117; blue, 5 }  ,draw opacity=1 ]   (170.25,176.5) -- (336.5,176.25) ;
\draw [color={rgb, 255:red, 65; green, 117; blue, 5 }  ,draw opacity=1 ]   (336.5,176.25) -- (336.5,296.75) ;
\draw  [color={rgb, 255:red, 208; green, 2; blue, 27 }  ,draw opacity=1 ] (475.5,281.75) .. controls (475.53,277.08) and (473.22,274.73) .. (468.55,274.7) -- (416.93,274.32) .. controls (410.26,274.27) and (406.95,271.92) .. (406.98,267.25) .. controls (406.95,271.92) and (403.6,274.22) .. (396.93,274.17)(399.93,274.2) -- (345.3,273.8) .. controls (340.63,273.76) and (338.28,276.07) .. (338.25,280.74) ;
\draw  [color={rgb, 255:red, 65; green, 117; blue, 5 }  ,draw opacity=1 ] (241,174.25) .. controls (244.29,174.32) and (245.97,172.7) .. (246.04,169.41) -- (246.04,169.41) .. controls (246.14,164.7) and (247.84,162.39) .. (251.13,162.46) .. controls (247.84,162.39) and (246.24,160) .. (246.34,155.29)(246.29,157.41) -- (246.34,155.29) .. controls (246.41,152) and (244.79,150.32) .. (241.5,150.25) ;

\draw (209.5,301.9) node [anchor=north west][inner sep=0.75pt]  [font=\small]  {$\delta t_{j}$};
\draw (533,291.4) node [anchor=north west][inner sep=0.75pt]    {$\delta t$};
\draw (154,25.4) node [anchor=north west][inner sep=0.75pt]    {$\Psi $};
\draw (143,214.9) node [anchor=north west][inner sep=0.75pt]  [font=\footnotesize]  {$\Psi _{j} \ $};
\draw (119,138.4) node [anchor=north west][inner sep=0.75pt]  [font=\footnotesize]  {$\Psi _{j} +\psi _{j} \ $};
\draw (125.5,66.9) node [anchor=north west][inner sep=0.75pt]  [font=\footnotesize]  {$\textcolor[rgb]{0.82,0.01,0.11}{\bar{\Psi}_{j}}\textcolor[rgb]{0.82,0.01,0.11}{+}\textcolor[rgb]{0.82,0.01,0.11}{\bar{\psi}_{j}}\textcolor[rgb]{0.82,0.01,0.11}{\ }$};
\draw (311.5,254.9) node [anchor=north west][inner sep=0.75pt]    {$\tau _{j}^{( i)}$};
\draw (190.5,180.4) node [anchor=north west][inner sep=0.75pt]  [font=\small]  {$\psi _{j} \ $};
\draw (421.5,271.5) node [anchor=north west][inner sep=0.75pt]  [color={rgb, 255:red, 65; green, 117; blue, 5 }  ,opacity=1 ,rotate=-270] [align=left] {{\scriptsize old putative infection time}};
\draw (465,255.5) node [anchor=north west][inner sep=0.75pt]  [font=\scriptsize,color={rgb, 255:red, 208; green, 2; blue, 27 }  ,opacity=1 ,rotate=-270] [align=left] {new putative infection time\\};
\draw (398,303.4) node [anchor=north west][inner sep=0.75pt]  [font=\footnotesize]  {$\delta t_{j} +\tau _{j}^{( i)}$};
\draw (220.5,218) node [anchor=north west][inner sep=0.75pt]  [color={rgb, 255:red, 0; green, 0; blue, 0 }  ,opacity=1 ,rotate=-90] [align=left] {{\scriptsize old current time}};
\draw (460.5,302.9) node [anchor=north west][inner sep=0.75pt]  [font=\footnotesize]  {$\textcolor[rgb]{0.82,0.01,0.11}{\bar{\delta} t_{j}}\textcolor[rgb]{0.82,0.01,0.11}{+}\textcolor[rgb]{0.82,0.01,0.11}{\bar{\tau}_{j}^{(i)}}$};
\draw (303.5,321.4) node [anchor=north west][inner sep=0.75pt]  [font=\small]  {$S=s$};
\draw (257.5,346.4) node [anchor=north west][inner sep=0.75pt]  [font=\small,color={rgb, 255:red, 65; green, 117; blue, 5 }  ,opacity=1 ]  {$S=s$};
\draw (389.5,346.9) node [anchor=north west][inner sep=0.75pt]  [font=\small,color={rgb, 255:red, 208; green, 2; blue, 27 }  ,opacity=1 ]  {$S=\bar{s}$};
\draw (264,217.5) node [anchor=north west][inner sep=0.75pt]  [font=\scriptsize,color={rgb, 255:red, 65; green, 117; blue, 5 }  ,opacity=1 ] [align=left] {update};
\draw (377.5,102) node [anchor=north west][inner sep=0.75pt]  [font=\scriptsize,color={rgb, 255:red, 208; green, 2; blue, 27 }  ,opacity=1 ] [align=left] {rescale};
\draw (299.5,94.9) node [anchor=north west][inner sep=0.75pt]  [font=\scriptsize]  {$\textcolor[rgb]{0.82,0.01,0.11}{\bar{\Psi}_{j}}\textcolor[rgb]{0.82,0.01,0.11}{=}\textcolor[rgb]{0.82,0.01,0.11}{\frac{\textcolor[rgb]{0.82,0.01,0.11}{s}}{\bar{s}}}\textcolor[rgb]{0.82,0.01,0.11}{\psi _{j}^{'} \ }$};
\draw (187,164.9) node [anchor=north west][inner sep=0.75pt]  [font=\scriptsize,color={rgb, 255:red, 65; green, 117; blue, 5 }  ,opacity=1 ]  {new $\Psi_{j}$};
\draw (187,205.4) node [anchor=north west][inner sep=0.75pt]  [font=\scriptsize]  {old $\Psi_{j}$};
\draw (186.5,133.4) node [anchor=north west][inner sep=0.75pt]  [font=\scriptsize]  {old $\Psi_{j} +\psi _{j}$};
\draw (184,78.4) node [anchor=north west][inner sep=0.75pt]  [font=\scriptsize]  {new $\Psi_{j} +\psi_{j}$};
\draw (347.5,188.5) node [anchor=north west][inner sep=0.75pt]  [color={rgb, 255:red, 65; green, 117; blue, 5 }  ,opacity=1 ,rotate=-90] [align=left] {{\scriptsize new current time}};
\draw (272.5,236.4) node [anchor=north west][inner sep=0.75pt]    {$\tau ^{( m)}$};
\draw (326.5,299.9) node [anchor=north west][inner sep=0.75pt]  [font=\footnotesize]  {$\textcolor[rgb]{0.25,0.46,0.02}{\bar{\delta}t_{j}}$};
\draw (389.5,243.9) node [anchor=north west][inner sep=0.75pt]    {$\bar{\tau}_{j}^{( i)}$};
\draw (143,168.9) node [anchor=north west][inner sep=0.75pt]  [font=\footnotesize]  {$\textcolor[rgb]{0.25,0.46,0.02}{\bar{\Psi}_{j}}$};
\draw (253.5,152.4) node [anchor=north west][inner sep=0.75pt]  [font=\small]  {$\psi _{j}^{'} \ $};
\end{tikzpicture}
\caption{\textbf{Schematic describing the process of putative infection time rescaling for a given individual $j$.}~The blue curve corresponds to the tabulated function $\Psi(\delta t)$ described by (\ref{nu}) and is fixed for all individuals.~The green components indicate updating to the current time within the infection process for this individual and the red components indicate the rescaling process for the next infection time.~All characters without bars indicate original parameter values, dashes indicate that time has been updated but scaling has not yet taken place, and overbars indicate that they are fully updated values.}\label{rescaleImage}
\end{figure}

\subsubsection{Implementation of accurate multiscale simulation}
\label{multiscale algo implementation}
To simulate the within-host and population-level integrated model, we take the framework of a typical population-level Gillespie-style NRM.~The algorithm updates from event to event by storing all putative times for each unique event and implementing the event corresponding to the most imminent.~Putative times for all Markovian events (in our case events ${(b)}$, ${(ds)}$, and ${(di)}$) are drawn in the same way as described in Section \ref{chap:SSA}.~However, times for the non-Markovian events (in our case ${(i)}$) need to be treated differently. We shall store in $\vec{\tau}$ a complete list of putative times for all Markovian events. In particular, $\vec{\tau} = \left( \tau^{(b)}, \tau^{(ds)}, \tau^{(di)} \right)^\intercal$. Furthermore, for each infected individual we also keep track of four pieces of information $\vec{\Psi} = \left\{\Psi_j \right\}_{j=1}^i$, $\vec{\psi} = \left\{\psi_j \right\}_{j=1}^i$, $\vec{\tau}^{(i)} = \left\{\tau_j^{(i)}\right\}_{j=1}^i$ and $\vec{\delta t} = \left\{\delta t_j \right\}_{j=1}^i$. Whilst it is the case that elements in $\vec{\Psi}$ are related to elements in $\vec{\delta t}$ in the same way as elements of $\vec{\Psi}+\vec{\psi}$ to those of $\vec{\delta t} + \vec{\tau}^{(i)}$, by the function $\Psi(\delta t)$, we find it convenient to store and track all four numbers for each infected individual.\\ 

To implement our method, we first obtain a numerical solution for the within-host model and, following (23), store $\Psi$ in a table.~For efficiency, as we explain later, we instead construct two look-up tables.~The first table, $\Psi(\delta t)$, is stored using a uniform discretisation of $\delta t$ with step size $\delta$, while the second table, $\delta t(\Psi)$, has the same length as $\Psi(\delta t)$ but is discretised with respect to $\Psi$ using step size $\delta \Psi$.~Importantly, whilst this approach is accurate, the primary source of error stems from the resolution of these tables, necessitating sufficiently small $\delta$ and $\delta \Psi$.~Then, to evaluate $\Psi$ for an infectious individual $j$ with age of infection $\delta t_j$, we retrieve its value from $\Psi(\delta t)$ by indexing the table at $\delta t_j / \delta$, rounded to the nearest integer.~Similarly, evaluating $\Psi^{-1}$ follows the same principle by accessing $\delta t(\Psi)$ at $\psi_j / \delta \Psi$, also rounded to the nearest integer.~From now on, we denote these look-up operations as $\Psi$ and $\Psi^{-1}$, respectively.~When applied to a vector, these operations are performed element-wise, yielding a corresponding vector of output values.~A key advantage of this approach is its computational efficiency.~Both $\Psi$ and $\Psi^{-1}$ operations involve only floating-point division and array indexing, which are computationally inexpensive, regardless of the resolution of $\delta$ and $\delta \Psi$ or the size of the tables $\Psi(\delta t)$ and $\delta t(\Psi)$.~This highlights a crucial benefit of our novel exact algorithm -- high efficiency is achieved alongside accuracy.\\

We begin by initialising all population compartments ($s = s_0$, $i = i_0$). For each infected individual $j=1,\ldots,i_0$, it is necessary to determine an initial calendar time of infection $t_j$ and use this time to initialise $\vec{\delta t}$. We use the look-up tables to initialise $\vec{\Psi} = \Psi(\vec{\delta t})$.~Next we initialise all putative times. We initialise $\vec{\tau}$ for Markovian events ($(b)$, $(ds)$, $(di)$) by finding respective propensities for each element and sampling putative times using (\ref{eventsampletime}).~We then initialise $\vec{\psi}$ for each element by using (\ref{xi}) and then use the look-up tables to initialise $\vec{\tau}^{(i)} = \Psi^{-1}(\vec{\Psi} + \vec{\psi}) - \vec{\delta t}$ for the non-Markovian event $(i)$.~The algorithm propagates forward by finding the most imminent event time $\tau^{(m)} = \mathrm{min}(\left\{\vec{\tau},\vec{\tau}^{(i)} \right\})$.~After finding the event, time is updated and then the event $(m)$ is implemented (changing the compartments and deleting or initialising elements of $\vec{\Psi}$, $\vec{\psi}$, $\vec{\delta t}$ and $\vec{\tau}^{(i)}$, respectively) and then its putative time is re-sampled whilst all other affected putative times should be re-scaled (either using (\ref{rescaling}) if it is a Markovian event or (\ref{rescalingpsi}) if it is a non-Markovian -- infection -- event). In particular, re-sampling and re-scaling of the non-Markovian infection events are described in Section \ref{clever} and involve updating respective elements of $\vec{\delta t}$, then $\vec{\Psi}$ followed by $\vec{\psi}$ and then $\vec{\tau}^{(i)}$.~The algorithm then propagates until some pre-determined final time.~As shorthand, we shall denote $\mathcal{Q} = \left\{\vec{\delta t}, \vec{\tau}^{(i)}, \vec{\Psi} , \vec{\psi} \right\}$ as the collection of vectors that store information about each infected individual.\\

The pseudocode for this algorithm is shown in Algorithm \ref{Efficient multiscale algorithm}, where the MATLAB implementation and all data generated from the simulations are available on Github at \url{https://github.com/YuanYIN99/Accurate-stochastic-simulation-algorithm-for-multiscale-models-of-infectious-diseases.git}.

\begin{algorithm}
  \caption{Accurate multiscale algorithm (omitting units for readability)}
  \begin{algorithmic}[1]
    \State \textbf{Input}: \\
    \quad Within-host model (for example (\ref{WH ODE})) to generate two look-up tables $\Psi(\delta t)$ and $\delta t (\Psi)$ as described in Section \ref{multiscale algo implementation};\\
    \quad Population-level parameters (for example, $\Lambda$ and $\mu$ defined by (\ref{BH submodel}));\\
        \quad Initial conditions for the population-level model, $t = 0$, and e.g. $\left(s(t), i(t)\right)=\left(s_0, i_0\right)$;\\
    \quad Initial conditions for the within-host model of each infected individual $\vec{\delta t}$ accompanied by the initial values of $\vec{\Psi} = \Psi(\vec{\delta t})$; \\
    \quad $t_{end}$: ending time for the simulation.\\
    \State Initialise all putative times for Markovian events $\vec{\tau}$ using (\ref{eventsampletime}).
    \State Initialise the putative $\vec{\psi}$ and times $\vec{\tau}^{(i)}$ for non-Markovian events using (\ref{xi}) and (\ref{tauupdate}).
    \State
    \While{$t \leq t_{end}$}
      \State Find the event $(m)$ and time increment $\tau^{(m)}$ corresponding to the minimum putative time $\tau^{(m)} = \mathrm{min}(\left\{\vec{\tau},\vec{\tau}^{(i)} \right\})$.
      \State Increment time forward to the new event; $t := t + \tau^{(m)}$, $\vec{\tau}:= \vec{\tau} - \tau^{(m)}$, $\vec{\delta t} := \vec{\delta t} + \tau^{(m)}$.
      \State Make appropriate changes to the number in each state (e.g. compartment copy numbers) due to event $(m)$.
      \State
      \If{the event $(m)$ changes the copy number of non-Markovian states (for example the number of infected individuals)}
      \State Remove or add corresponding individual elements from $\mathcal{Q}$ as appropriately defined by $(m)$. To add a new element to the vectors in $\mathcal{Q}$, begin by setting the element $\delta t_j = 0$, $\Psi_j = 0$ and sampling $\psi_j$ using (\ref{xi}) and $\tau_j^{(i)} = \Psi^{-1}(\psi_j)$.
      \EndIf
      \State
      \If{event $(m)$ is Markovian} resample the putative time for event $(m)$ using (\ref{eventsampletime}). 
      \Else resample non-Markovian putative time for event $(m)$ using the method in Section \ref{clever}.
      \EndIf
      \State
      \State For each event Markovian event $(e)\neq(m)$, rescale the putative times in $\vec{\tau}$ using (\ref{rescaling}).
      \State For each individual $j$ such that $\tau_j^{(i)}$ does not correspond to $(m)$, \If{the propensity is affected by the change of state} 
      \State update $\Psi_j$, rescale $\psi_j$, and then calculate the rescaled $\tau_j^{(i)}$ as described in Section \ref{clever}. \Else
      \State update the putative time due to the time update $\tau_j^{(i)} := \tau_j^{(i)} - \tau^{(m)}$. \EndIf
    \EndWhile\\
    \State \textbf{Return} $s$ and $i = \mathrm{dim}(\vec{\delta t})$ after each event along with the times $\tau^{(m)}$ of those events.
  \end{algorithmic}
  \label{Efficient multiscale algorithm}
\end{algorithm}

\newpage
\section{Results}\label{results}
We now evaluate the performance of our novel algorithm by comparing its accuracy and efficiency against the time-driven approximate method.~We focus on a test case defined by the within-host dynamics in Table \ref{tab:WH param and var} and the population-level parameters in Table \ref{tab:BH param and var}.~In this section, we investigate how varying the initial population sizes ($(S, I) = (100, 10)$, $(200, 20)$, $(400, 40)$, and $(800, 80)$ people) and the frequencies at which within-host information is harvested affects the performance of both algorithms.~Each simulation is performed with $800$ repetitions using parallel computing, and all simulations are executed on a local computer with a 14-core CPU. 
\begingroup
\centering
\begin{table*}[]
\begin{center}
\begin{tabular}{m{4cm} m{4cm}}
\hline
\hline
\textbf{Initial conditions} & \textbf{Values for $I$}\\
\hline
$C_0$ & $10^8$ health cells\\
\hline
$C_0^*$ & $10^0$ infected cells\\
\hline
$V_0$ & $10^6$ viral particles\\
\hline
\hline
\textbf{Parameters} & \textbf{Values}\\
\hline
$p$ & $10 \ \text{day}^{-1}$ \\
\hline
$c$ & $1 \ \text{day}^{-1}$\\
\hline
$k$ & $10^{-7} \ \text{day}^{-1} \ \text{cell}^{-1}$\\
\hline
$\mu_c$ & $10^{-1} \ \text{day}^{-1}$\\
\hline
$\delta_c$ & $10 \ \text{day}^{-1}$\\
\hline
$\Lambda_c$ & $(1.111 \times 10^{6}) \ \text{cell} \times \text{day}^{-1}$\\
\hline
\hline
\end{tabular}
\end{center}
\caption{\textbf{Within-host initial conditions and parameter values used in Section \ref{results}.} Note that for the susceptible population, their within-host initial conditions are set as $(C_0,C_0^*, V) = (10^8, 0, 0)$ number of cells or viral particles instead.}
\label{tab:WH param and var}
\end{table*}
\endgroup
\begingroup
\centering
\begin{table*}[]
\begin{center}
\begin{tabular}{m{2.2cm} m{5.8cm}}
\hline
\hline
\textbf{Parameters} & \textbf{Values}\\
\hline
$\Lambda$ & $10^{-5}\text{ person}\times \text{day}^{-1}$ \\
\hline
$\mu$ & $10^{-5}\rm \text{ }day^{-1}$\\
\hline
$l$ & $10^{-9.8} \text{ }\rm viral \text{ } particle^{-1} \text{ }day^{-1}\rm person^{-1}$\\
\hline
\hline
\end{tabular}
\end{center}
\caption{\textbf{Population-level and the linking parameter values used in Section \ref{results}.}}
\label{tab:BH param and var}
\end{table*}
\endgroup
\subsection{Accuracy}
\label{sec: accuracy}
\begin{figure}[t!]
    \centering
\includegraphics[width=0.7\textwidth]{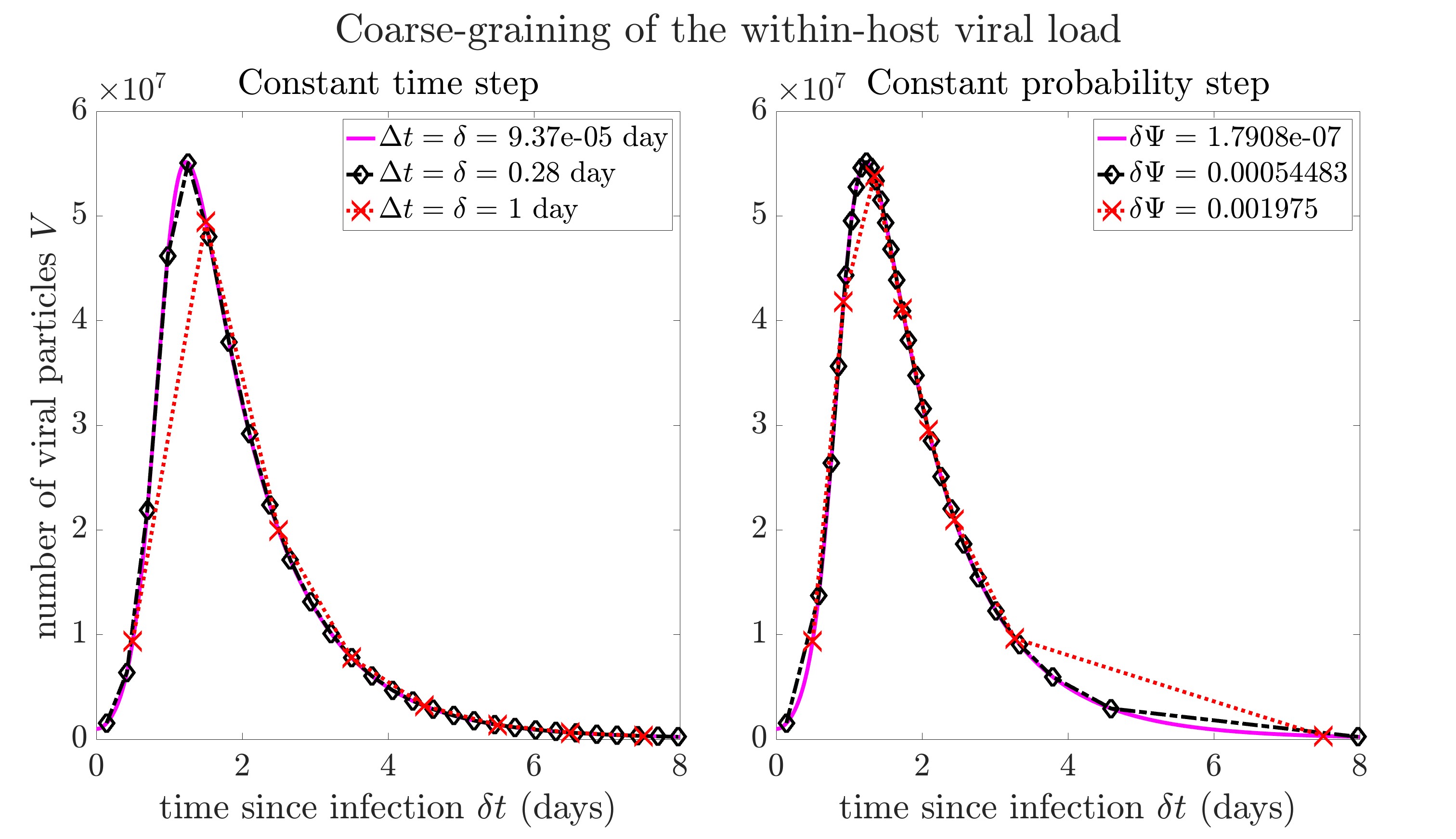} 
    \caption{\textbf{The resolution of harvesting the within-host viral load using a constant time step $\Delta t = \delta $ and a constant probability step $\delta \Psi$.}~Note that $\delta \Psi$ is defined so that the within-host system is extracted at the same resolution as $\delta$.~The within-host system, introduced in (\ref{WH ODE}) in Section \ref{chap:WH}, is solved using MATLAB's $ode15s$ solver with both absolute and relative tolerances set to $10^{-8}$.~The initial conditions and the parameter values for the within-host system are listed in Table \ref{tab:WH param and var}.}
    \label{fig:WH extraction}
\end{figure}

\begin{figure}[tbh]
    \centering
    \includegraphics[width=\textwidth]{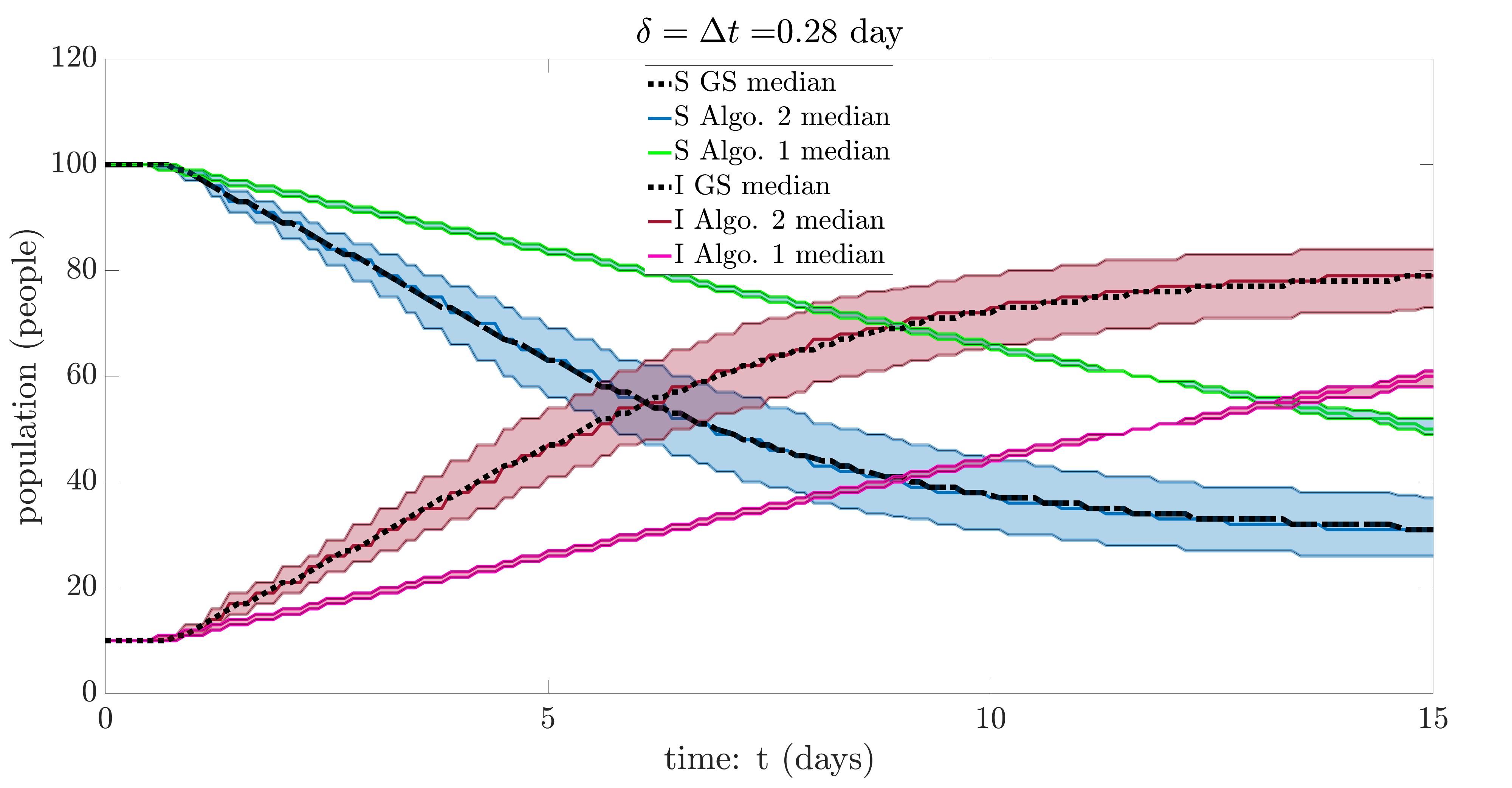} 
    \caption{\textbf{Population-level $SI$ dynamics with $\Delta t = \delta  = 0.28$ day.}~The highlighted bands enclose the 25th to 75th percentiles.~We set the population-level initial condition $(S, I) = (100, 10)$ people and simulate 800 runs.~The initial conditions and parameter values for the within-host system are listed in Table \ref{tab:WH param and var}.~The parameter values chosen for the population-level model are listed in Table \ref{tab:BH param and var}.}
    \label{fig:pop dynamics}
\end{figure}

\begin{figure}[tbh]
    \centering
    \includegraphics[width=0.8\textwidth]{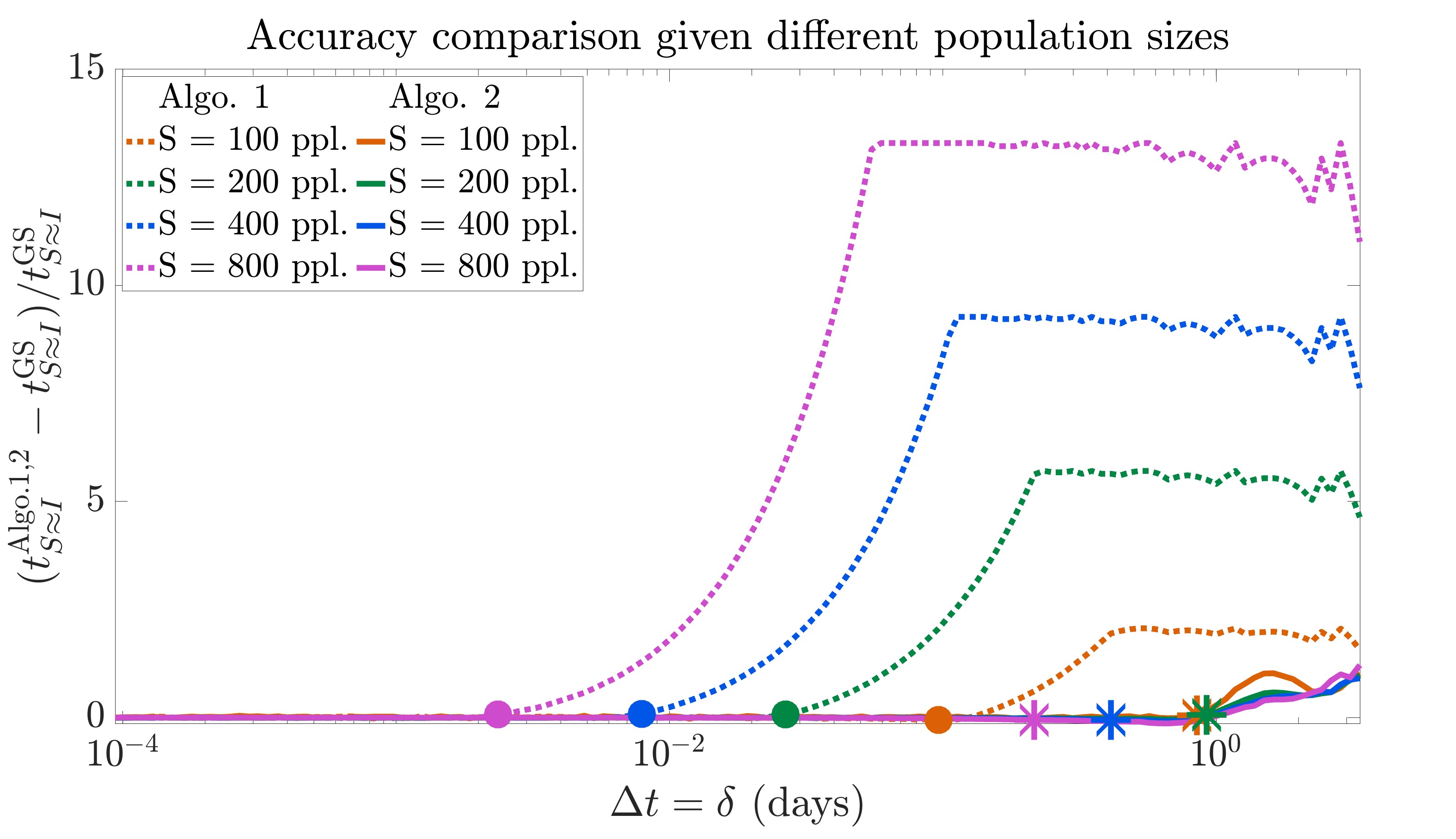} 
    \caption{\textbf{Accuracy for the approximate Algorithm \ref{Approximate time-driven algorithm} and the novel Algorithm \ref{Efficient multiscale algorithm} compared with the GS}. This plot displays the relative errors in the time point $t_{S\approx I}$, where the susceptible population equals or best approximates the infectious population, with different initial conditions $(S, I)=(100, 10)$, $(200, 20)$, $(400, 40)$, and $(800, 80)$ people, averaged over 800 repetitions.~The solid `dot' markers for the approximate time-driven algorithm represent the maximal $\Delta t_{\rm crit}$ such that for all $\Delta t \geq \Delta t_{\rm crit}$, the relative error is no smaller than $0.05$.~The `asterisk' markers serve a similar purpose for the novel algorithm.~The initial conditions and parameter values for the within-host system are listed in Table  \ref{tab:WH param and var}.~The parameter values chosen for the population-level model are listed in Table \ref{tab:BH param and var}.}
    \label{fig:accuracy plot}
\end{figure}
In this section, we aim to validate the accuracy of our novel algorithm (Algorithm \ref{Efficient multiscale algorithm}) compared to the approximate time-driven method (Algorithm \ref{Approximate time-driven algorithm}).~In the approximate Algorithm \ref{Approximate time-driven algorithm}, the calendar time is discretised with finite fixed time step $\Delta t$, while in our novel Algorithm \ref{Efficient multiscale algorithm}, time discretisation is limited to the resolution of tabulated $\Psi(\delta t)$ thus $\delta t(\Psi)$ associated with the within-host model.~Recall that since the length of the latter look-up table $\delta t(\Psi)$ is equal to that of the former $\Psi(\delta t)$, the discretisation of $\Psi$ in the inverse table is determined by the time discretisation $\delta$.~We therefore characterise the discretisation of both look-up tables in the novel Algorithm \ref{Efficient multiscale algorithm} as $\delta$.~Note that it is possible to make $\delta$ very small without incurring the cost of reduced efficiency in using these tables (see Section \ref{multiscale algo implementation}).~It's worth noting that whilst $\delta$ represents a coarse-graining of the within-host model dynamics, the calendar time discretisation $\Delta t$ represents a course-graining in within-host and population scale models.~That is, it is possible to miss intra-timestep population scale events with Algorithm \ref{Approximate time-driven algorithm} but not with Algorithm \ref{Efficient multiscale algorithm}.~However, if $\Delta t \rightarrow 0$, Algorithm \ref{Approximate time-driven algorithm} is expected to yield accurate results.~Numerically, the $\Delta t_{\rm min}$ which guarantees the accuracy of the approximate algorithm is chosen such that
$$\max_{\Delta t}\left\{\alpha^{(b)} + \alpha^{(ds)} + \alpha^{(di)} + s\sum_{j = 1}^i v(t-t_j)\right\}\Delta t \leq \epsilon:= 0.01,$$ 
where $\max_{\Delta t}\left\{s\sum_{j = 1}^i v(t-t_j)\right\} \leq \max_{\Delta t}\{v\} \times(s+i)^2/4$.~For the initial condition $(S, I) = (800, 80)$ people, we have $\Delta t_{\min} = 9.37 \times 10^{-5}$ day based on parameter values and within-host initial conditions shown in Tables \ref{tab:WH param and var}-\ref{tab:BH param and var}.~To assess the accuracy of our novel algorithm, we contrast consistent discretisation $\delta = \Delta t$ between the two algorithms and vary this discretisation geometrically over $131$ points.~Specifically, $\delta = \Delta t = \Delta t_{\min} \times 1.084^{(0:130)}$ days.\\

In \autoref{fig:WH extraction}, we visualise the coarse-graining of the within-host viral load $V$ using constant time steps $\delta = 9.37 \times 10^{-5}$ days, $\delta = 0.28$ days, and $\delta = 1$ day, along with the corresponding constant probability steps, assuming that the within-host virus dynamics follow the model parameters specified in Table \ref{tab:WH param and var}.~Recall that constant time tabulation is used in the time-driven Algorithm \ref{Approximate time-driven algorithm} as well as in the computation of $\Psi(\delta t)$ in our novel exact Algorithm \ref{Efficient multiscale algorithm}, whereas constant probability tabulation is employed for computing $\Psi^{-1}$ in Algorithm \ref{Efficient multiscale algorithm}.~We assume that an infection begins on a time step and thus we sample halfway between time steps as an approximation of the viral load on this time interval.~We refer to the error associated with the discretisation of the within-host model as `resolution error'.~We are not deterred from referring to Algorithm \ref{Efficient multiscale algorithm} as an exact algorithm, since it is, at least in principle, exact -- provided that the continuous function $\Psi(\delta t)$ (and consequently $\Psi^{-1}$) is known -- just as the Gillespie algorithm is considered exact despite potential numerical errors in the evaluation of propensities.\\

We designate the approximate time-driven algorithm with $\Delta t_{\min} = 9.37 \times 10^{-5}$ day as the ‘Golden Standard’ (GS), against which we will compare our novel algorithm.~\autoref{fig:pop dynamics} compares the population-level dynamics of the approximate Algorithm \ref{Approximate time-driven algorithm} and our novel Algorithm \ref{Efficient multiscale algorithm} with the GS, given $\Delta t = \delta  = 0.28$ day.~The `diamond' markers in \autoref{fig:WH extraction} visualise how we extract the within-host dynamics based on such resolution.~One can observe that as long as the within-host dynamics is approximated at a satisfactory level, or equivalently the `resolution error' is not surprisingly dominant, our novel algorithm converges to the GS and is indeed accurate.~In contrast, the dynamics of the approximate time-driven algorithm are notably slower than those of the GS. This is because the time-driven Algorithm \ref{Approximate time-driven algorithm} suffers from inaccurate approximation of probabilities for state changes at the population-scale which are not present in Algorithm \ref{Efficient multiscale algorithm}. \\

We now quantify the accuracy comparison between the two algorithms.~To quantify the error produced by the algorithm, we determine the calendar time in the test model at which the infectious and susceptible populations are equal $t_{S\approx I}$.~We then subtract this time from the time in the GS simulation and then divide the result by the GS time to get a relative error.~This error is plotted in ~\autoref{fig:accuracy plot} against the choice of $\Delta t = \delta$.~The solid curves show the error associated with Algorithm \ref{Efficient multiscale algorithm} whilst dotted curves present the error associated with Algorithm \ref{Approximate time-driven algorithm}.~The different colours indicate the effect of population size. In particular we use the different initial conditions: $(S, I) = (100, 10)$, $(200, 20)$, $(400, 40)$, and $(800, 80)$ people, respectively.~The coloured `dot' markers indicate the largest time step $\Delta t_{\mathrm{crit}}$ possible in Algorithm \ref{Approximate time-driven algorithm} in order that the error is kept below $5\%$, for comparison the coloured `asterisk' markers indicate the same limitation on $\delta_{\mathrm{crit}}$ in Algorithm \ref{Efficient multiscale algorithm}.~It can be observed that as we scale up the population, $\Delta t_{\mathrm{crit}}$ decreases linearly in the log scale of~\autoref{fig:accuracy plot}, whereas $\delta_{\mathrm{crit}}$ is relatively unaffected by population size and good accuracy can be achieved with relatively coarse discretisations of around $\delta = 0.1$ day.~The error observed for Algorithm \ref{Efficient multiscale algorithm} is due solely to `resolution error' which is a discretisation of within-host model but will not affect the population-scale, as previously explained.

\subsection{Efficiency}
\begin{figure}[t!]
    \centering
    \includegraphics[width=0.8\textwidth]{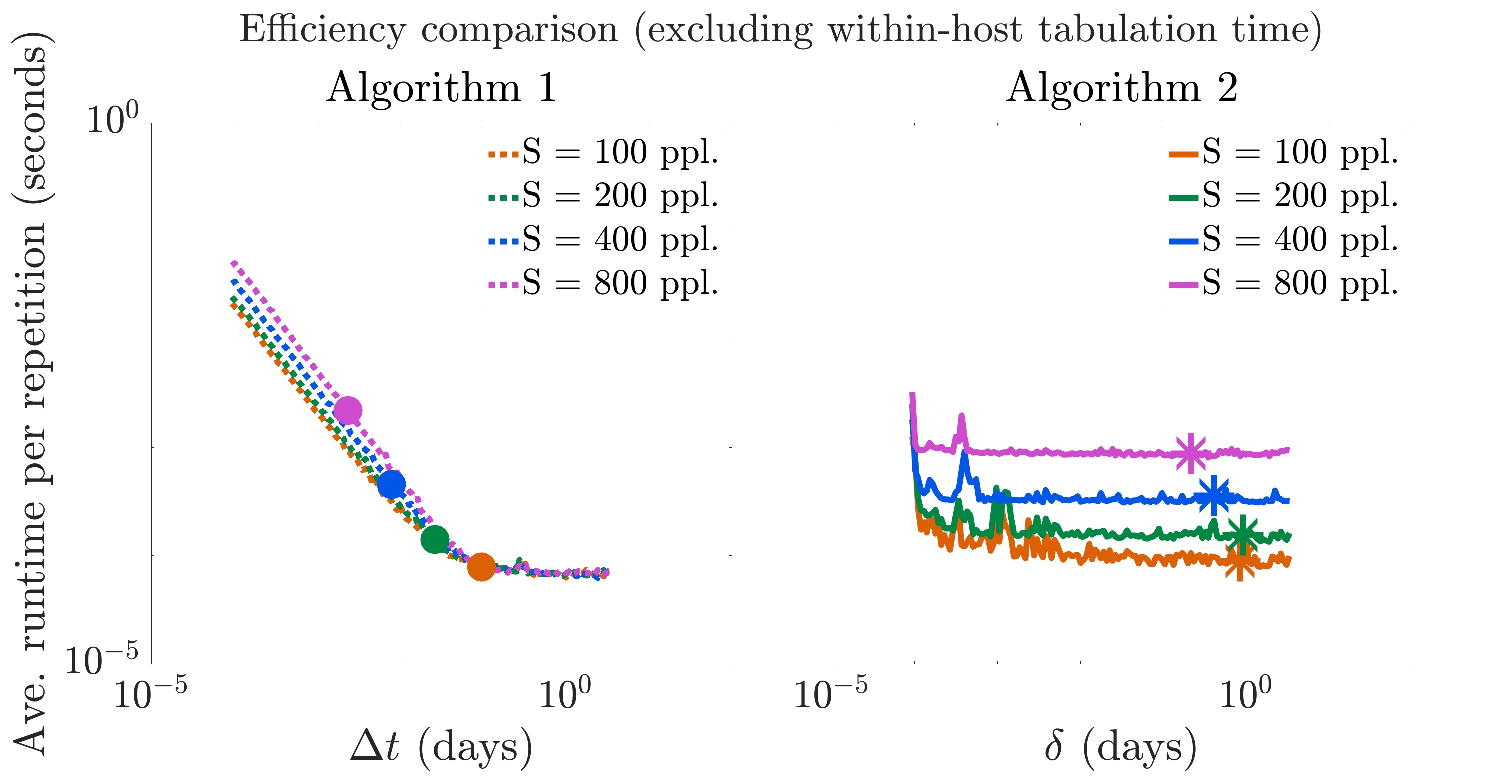} 
    \caption{\textbf{Average runtime (excluding within-host tabulation time) of the approximate time-driven Algorithm \ref{Approximate time-driven algorithm} and the novel Algorithm \ref{Efficient multiscale algorithm} for simulations shown in Figure \ref{fig:accuracy plot}}.~The orange, green, blue, and light purple curves represent different initial conditions: $(S, I)=(100, 10)$, $(200, 20)$, $(400, 40)$, and $(800, 80)$ people, respectively.~The markers indicate the maximum value of $\delta = \Delta t$ that generates accurate results, with a relative error bounded by $0.05$.~The initial conditions and parameter values for the within-host system are provided in Table \ref{tab:WH param and var}, while the population-level model parameter values are listed in Table \ref{tab:BH param and var}.}
    \label{fig:efficiency plot-noWH}
\end{figure}

In this section, we aim to validate the efficiency of our novel algorithm (Algorithm \ref{Efficient multiscale algorithm}) compared to the time-driven method (Algorithm \ref{Approximate time-driven algorithm}).~The primary computational cost in the approximate time-driven algorithm arises from tabulating the force of infection $v$ at the start of the simulation, which has a complexity of $O(1/\Delta t)$, and determining the next event.~For the latter, given a calendar time discretisation of $\Delta t$, we must sum over $I$ forces of infection at each time step, leading to a time complexity of $O(I/\Delta t)$.~Consequently, the overall complexity of the approximate time-driven algorithm is $O(I/\Delta t)$.~In contrast, the computational complexity of the novel exact algorithm primarily stems from two factors:~(1) constructing two look-up tables of resolution $\delta$ at the start of the simulation and (2) determining event-driven times while updating the within-host systems of infectious individuals. Construction of the two look-up tables has a complexity of $O(1/\delta)$, while the latter operation is $O(I)$, as it only involves floating-point divisions and array indexing for each infectious individual.~If a total of $M$ events occur before the simulation end time, $t_{\text{end}}$, the overall time complexity of our exact algorithm is $O(MI + 1/\delta)$.~An important observation is that, due to the rejection nature of the time-driven algorithm, the number of events satisfies $M\leq t_{\text{end}}/\Delta t$.~Therefore, with resolutions $\Delta t = \delta$, the time-driven algorithm exhibits no better time complexity compared to our novel exact algorithm, when the tabulation time is excluded (as it's a fixed cost).\\

We then compare the efficiency of the novel exact Algorithm \ref{Efficient multiscale algorithm} with the approximate time-driven Algorithm \ref{Approximate time-driven algorithm}, across varying population sizes and resolutions for extracting within-host information. Figure \ref{fig:efficiency plot-noWH} shows the average runtime per repetition, excluding the time required to tabulate the within-host system.~We observe that the runtime of Algorithm \ref{Approximate time-driven algorithm} increases with finer time steps ($\Delta t$) and bigger population sizes.~This observation is consistent with the previous theoretical complexity analysis.~On the other hand, the runtime of Algorithm \ref{Efficient multiscale algorithm} is independent of the resolution used to tabulate within-host information, but increases with population size, reflecting the greater number of events occurring over the simulation time frame.~Additionally, the runtime of Algorithm \ref{Efficient multiscale algorithm} appears less smooth due to stochastic variability in the number of events per simulation.~We now focus on regions where the time discretisations ($\Delta t$ and $\delta$) are smaller than those highlighted by the markers in Figure \ref{fig:efficiency plot-noWH}, as these indicate the critical resolutions required to ensure accuracy (see Section \ref{sec: accuracy}).~In practice, we find that the novel exact algorithm is more efficient -- or at least equivalent -- to the time-driven approach in these regimes.~We also note that, in this specific test case, the tabulation time dominates the total runtime.~However, this need not be the case in other scenarios -- for example, when the within-host system stabilises rapidly.

\section{Discussion}\label{discussion}
When simulating a population-scale model of the spread of a disease, it is not uncommon to see the implementation of autonomy (memorylessness) in the model.~This is the case with ODE and also Gillespie-type stochastic simulations.~In small populations, where the outcome of an epidemic may hinge on the infections of one or two infected individuals, these memoryless models fail to account for the stages of their infection.~To track an individual's viral load, a within-host model is required to be coupled to the population-scale model.~We have shown that the introduction of a within-host model for each individual can be achieved with extremely small time steps.~However, such a time-driven algorithm becomes inaccurate (Figures \ref{fig:pop dynamics}-\ref{fig:accuracy plot}) if the fixed time step is no longer small enough.~It is important that the time step chosen is much smaller than the natural rate of change of viral load for an infected individual.~Additionally, since working with predefined finite time steps, it is important to keep the time step much smaller than the inverse of the rate of infection and therefore the time step should depend on the unknown future population-level and the overall within-host level dynamics, which is almost impossible to estimate.~This is because it is assumed that the probability for an event to occur in any given time step should be small.~Whilst Tau-leaping style approaches might also be useful, they too require some consideration for the rate at which events are occurring in order to retain sufficiently high accuracy.\\

In this paper, we have sidestepped this challenge and developed a novel stochastic algorithm (Algorithm \ref{Efficient multiscale algorithm}) for multiscale systems which incorporates a within-host viral load dynamic into the infection rates of individuals embedded in a population-scale model of an infectious disease.~The key strength of our algorithm lies in its accuracy, efficiency, and generality.~Specifically, event times are simulated in a similar way to the Gillespie algorithm but taking into consideration the explicit time dependence in individual viral loads.~We conclude from Section \ref{results} that as long as the within-host data are extracted at a sensible resolution, the simulated results will always be accurate using our approach.~Furthermore, extracting within-host information at finer resolutions does not compromise efficiency (Figure \ref{fig:efficiency plot-noWH}).~While we have introduced the concepts of this novel approach using simple target cell-limited within-host and $SI$ population-level models, our algorithm can be applied to similar compartmental models at both levels.~For example, for a different within-host model, the primary change would be in the look-up table $\Psi$ ((\ref{nu}) in Section \ref{MultiscaleSSA}), which records the within-host quantities that influence the population-level dynamics.~For a different population-level model, we would need to consider different events, such as infection, recovery, and waning immunity.~However, the method for sampling non-Markovian event putative times using the look-up table, resampling and rescaling those times after an event, remains unchanged.~Therefore, we conclude that our accurate Algorithm \ref{Efficient multiscale algorithm} can be broadly applied across the field of multiscale infectious disease modelling.\\

This novel algorithm can also be extended in several directions.~For instance, we can explore scenarios where the whole population still shares a single within-host system, but it is no longer deterministic and includes noise.~Alternatively, we can build a heterogeneous population.~For instance, in age-structured populations where each sub-population shares a single deterministic within-host system, one can simply create different look-up tables that tabulate different within-host dynamics for each sub-population.~Moreover, we can extend our algorithm to investigate situations where different infectious individuals have varying within-host time scales and amplitudes by tabulating a dimensionless version of the solution of the within-host model.~However, we emphasise that our algorithm cannot be extended to scenarios where each individual has a unique set of within-host system.~In such cases, it would be extremely challenging for any simulation approach to generalise.\\

Our work also holds broader implications and significance in real-world scenarios.~Until now, considerable effort has been dedicated to nesting the within-host system into the population-level one in multiscale modelling of infectious diseases \cite{tatsukawa2023agent, martcheva2015coupling, mideo2008linking}.~However, the reverse direction -- coupling population-level dynamics into the within-host system -- has not been fully explored \cite{feng2012model}.~A key step in investigating this reverse nesting is obtaining a global within-host system by averaging over the entire population.~The novel and accurate Algorithm \ref{Efficient multiscale algorithm} we have developed is an ideal tool for generating and extracting such information.~This allows for uncovering potential relationships between population-level dynamics and global within-host parameters, enabling the derivation of a single set of ODEs that incorporates dynamics at both the population and global within-host levels after coarse-graining.~As a result, it would be much more efficient to test, refine, and calibrate the simple yet comprehensive coarse-grained ODE model to real-life data.~Parameter inference would also be more realistic, enabling predictions and the formulation of policies to slow down disease transmission.

\section{Code and data accessibility}
Codes for Algorithms \ref{Approximate time-driven algorithm}-\ref{Efficient multiscale algorithm} and all data generated from the simulations are available on Github at \url{https://github.com/YuanYIN99/Accurate-stochastic-simulation-algorithm-for-multiscale-models-of-infectious-diseases.git}.

\section{Acknowledgements}
J.A. Flegg’s research is supported by the Australian Research Council (DP200100747, FT210100034) and the National Health and Medical Research Council (APP2019093).~Y.Yin is funded by the Engineering and Physical Sciences Research Council (EP/W524311/1).~This research was supported by The University of Melbourne’s Research Computing Services and the Petascale Campus Initiative.~All authors declare that the research was conducted in the absence of any commercial or financial relationships that could be construed as a potential conflict of interest.

\section{Supplementary Information}
\subsection{Linear stability analysis at the population level }
\label{SI:linear stability}
The SI model (\ref{BH submodel}) in Section \ref{population level model} has tractable fixed points $(S,I) = (S^*,I^*)$.~They are 
$$(S^*,I^*) = (S_0^*,I_0^*) = (\Lambda \mu^{-1}, 0) \text{ and }(S^*,I^*) = (S_1^*,I_1^*) = ( \mu \beta^{-1}, [\Lambda \beta - \mu^2] (\beta \mu )^{-1}).$$
We shall refer to $(S_0^*,I_0^*)$ as the elimination state and $(S_1^*,I_1^*)$ as the endemic state.~It is possible to show through linear stability and bifurcation analysis that a transcritical bifurcation occurs when $\Lambda \beta = \mu^2$.~If $\Lambda \beta > \mu^2$, then the disease becomes endemic with $(S_1^*,I_1^*)$ being positive and stable whilst the disease elimination state $(S_0^*,I_0^*)$ is unstable.~If $\Lambda \beta < \mu^2$, then $(S_1^*,I_1^*)$ has a negative infectious population and is therefore an irrelevant fixed point and $(S_0^*,I_0^*)$ is stable.\\

~As an ODE, the $SI$ model only describes large continuous populations, and when populations $S$ or $I$ are small, the model is inappropriate.~For example, if $I_0>0$ and $\Lambda \beta > \mu^2$, then $I$ may get arbitrary small.~However, since $(S_0^*,I_0^*)$ is unstable, disease elimination is impossible.~On the other hand, even if $\Lambda \beta > \mu^2$, elimination should be possible when $I$ is small.~Because in reality a small number of infected individuals may, stochastically, recover or die before infecting anybody.~Such an elimination event can be achieved by a stochastic simulation of model (\ref{BH submodel}).~For instance, a discrete and stochastic transition of $I=1$ individual to $I=0$ leads exactly to the elimination steady state $(S_0^*,I_0^*)$.~This is an absorbing state since small perturbations are not possible in the discrete stochastic (and thus realistic) scenario.~Therefore, it is important for us to extend the deterministic population-level $SI$ model to an SSA.

\clearpage
\bibliographystyle{alpha}
\bibliography{sample}

\end{document}